\documentclass[]{emulateapj}

\newcommand{\vkms}{km s$^{-1}$~}

\slugcomment{Accepted to AJ}

\begin{document}

\title{Do the nearby BHB stars belong to the Thick Disk or the Halo? }

\author{T.D.Kinman\altaffilmark{1}}          
\affil{NOAO, P.O.Box 26732, Tucson, Arizona 85726, USA}

\author{Morrison, Heather, L. }                             
\affil{Department of Astronomy, Case Western Reserve University, 
 Cleveland,OH 44106, USA}

\author{Brown, Warren, R. }                             
\affil{Smithsonian Astrophysical Observatory, 60 Garden St.,    
 Cambridge, MA 02138, USA}

\altaffiltext{1}{ 
 The NOAO are operated by AURA, Inc.\ under cooperative 
agreement with the National Science
Foundation.}

\begin{abstract}

We study the Milky Way region 
 ($\mid$Z$\mid <$3.0 kpc), where the thick disk and inner halo overlap, 
  by using the kinematics of local blue horizontal branch (BHB) stars 
(within 1 kpc) and new samples of BHB stars and A-type stars from the
{\em Century Survey}. We derive Galactic $U,V \& W$ velocities for these BHB and
A-type star samples using proper motions from the {\it NOMAD} catalog. The mean 
velocities and the velocity dispersions of the BHB samples 
($\mid$Z$\mid$$<$3 kpc) are characteristic 
of the halo, while those of the {\it Century Survey} A-type stars are
characteristic of the thick disk. There is no evidence from our samples that the
BHB stars rotate with the thick disk in the region $\mid$Z$\mid$$<$3 kpc. 
 Nearly a third of
the nearby local RR Lyrae stars have disk kinematics and are more metal-rich 
than [Fe/H]$\sim$$-$1. 
Only a few percent of the {\it Century Survey} BHB stars have these properties.
Only one nearby BHB star (HD 130201) is likely to be such a disk star but 
selection based on high proper motions 
will have tended to exclude such stars from the local sample. 
The scale height derived from a sample of local RR Lyrae stars agrees with that 
of the {\it Century Survey} BHB stars. The local samples of BHB stars and 
metal-weak red giants are too incomplete for a similar comparison.

\end{abstract}


\section{Introduction}

  The separation of the stellar thin disk, thick disk and halo populations near
  the Galactic plane is challenging because all three populations overlap
  spatially.
  The situation has become even more complex since Morrison et al. (2008) 
  identified a new inner halo component with a vertical scale height comparable to
  that of the thick disk, which has, however, quite different kinematics. 
  Inner halo stars have predominantly eccentric orbits, 
  a much greater velocity dispersion than thick disk stars and no rotation. 
  Thick disk stars, on the other hand, have kinematics that 
   are dominated by rotation.
  Even so, one cannot unambiguously assign stars to the thick disk or halo
  using kinematics alone; metallicity provides an additional clue.

  Almost all inner halo stars have [Fe/H] $<$ $-$0.8, with a mean [Fe/H] of $-$1.6,
  whereas the mean [Fe/H] of the thick disk is about $-$0.5.
 The extent of the metal-weak tail of the thick disk is still uncertain. 
 The first identifications of metal-weak thick disk stars were made using
 samples of red giants whose metallicity calibration was later shown to be
 unreliable (Norris et al., 1985; Morrison et al., 1990 and Twarog \& 
 Anthony-Twarog, 1994).  Current studies of halo samples that have well-determined
 [Fe/H] less than $-$1.0 (e.g Morrison et al., 2008) show almost no disk stars, 
 but they are based on surveys that avoid the Galactic plane. On the other hand,
 35\% of the local RR Lyrae stars have disk kinematics although
 only a few of these disk RR Lyrae stars are more metal-poor than [Fe/H] 
 = $-$1.0. This is discussed further in the Appendix.

  BHB  and RR Lyrae stars are among 
 the most-used probes of the Galactic halo because both types of stars are 
effectively ``standard candles". Local BHB stars (within 1 kpc) have been
 reliably identified by high-resolution 
 spectroscopy (Kinman et al. 2000; Behr 2003); their kinematics have not
 been discussed in detail, although their mean $V_{LSR}$ is like that of the halo
 (Kinman et al. 2007 (Table 10)). 
 Recently, Brown et al. (2008)
 in their {\it Century Survey Galactic Halo Project}, have identified 
 655 non-kinematically selected BHB stars. In this sample, those 
 with 5 $< \mid$Z$\mid <$ 9 kpc show halo kinematics with a mean Galactic 
 velocity ($V_{LSR}$) of $\sim -$220 \vkms. In the region 
  2$<$ $\mid$Z$\mid <$ 5.5 kpc, however, Brown et al. found a gradient 
in this velocity of d$V_{LSR}$/d$\mid$Z$\mid$  = $-$28 $\pm$ 3.4 \vkms.
  They also found a density scale height for the BHB stars of 1.26 $\pm$ 0.1 kpc 
 and concluded that the BHB stars near the plane belong to the thick disk with a
 local space density of 104 $\pm$ 37 kpc$^{-3}$. 

  Brown et al. determined the mean Galactic velocity ($<V>$)
   of their BHB stars from 
 radial velocities \emph{alone} since they considered that the existing proper 
 motions were not accurate enough for them to get reliable $U,V$ \& $W$ 
 velocities for each star. 
 While this is certainly true of the proper motions for the majority of the stars
 in their survey, it seems possible that the proper motions of their nearer 
 stars (within 3 kpc) may be accurate enough to derive $U$, $V$ \& $W$ 
 velocities that could  show whether these stars belong to the disk
 or to the halo. Our purpose is to examine this possibility.

\section{Galactic Kinematics and the separation of disk from halo BHB stars}
    
 The Galactic $U$, $V$ and $W$ were derived from the proper motions, radial velocities and
 distances of each star using the program of Johnson \& Soderblom (1987) that gives 
 heliocentric velocities in a right-handed coordinate system that is positive towards the
 Galactic center, the direction of rotation and the NGP. These heliocentric 
 velocities were converted to those relative to the local standard of rest (LSR)
 $U_{LSR}$, $V_{LSR}$ and $W_{LSR}$ using the solar motion $U_{\odot}$ = +10 
 \vkms, $V_{\odot}$ = +5 \vkms and
 $W_{\odot}$ = +7 \vkms (Dehnen \& Binney, 1998). The proper motions were taken 
 from the \emph{NOMAD} catalog (Zacharias et al. 2004a) using the \emph{VizieR} 
  access tool.  This catalog lists the ``best" proper motion available
 for a particular star from the following catalogs: Hipparcos (ESA, 1997), Tycho-2 
 (Hog et al. 2000), UCAC2 (Zacharias et al. 2004b) and USNO-B-1.0 (Monet et al. 2003). 
 Most of the local sample have proper motions in the Hipparcos catalog and most of those in 
 the {\it Century Survey} come from the UCAC2 catalog; only 17\% of the proper motions of the
 {\it Century Survey} stars that are more distant than 2.75 kpc    
  come from the less accurate USNO-B-1.0 
  catalog. The few stars in the {\it Century Survey} samples whose proper 
 motions are not given in the \emph{NOMAD} catalog were not used.
 The sources of the radial velocities and distances are discussed
 separately for each sample. All velocities are in \vkms and we define the
    total space velocity ($T$) with respect to the LSR  as  
   ($V_{LSR}^{2}$ + $U_{LSR}^{2}$ +$W_{LSR}^{2}$)$^{0.5}$.

\begin{deluxetable*}{ cccccccccccc }
\tablewidth{0cm}
\tabletypesize{\footnotesize}
\setcounter{table}{1}
\tablecaption{ Comparison of the Galactic Kinematics of the various Samples.}
 
\tablehead{ 
\colhead{  Sample \tablenotemark{a} } &
\colhead{ N            \tablenotemark{b}  } &
\colhead{ $<U_{LSR}>$                     } &
\colhead{ $<V_{LSR}>$                     } &
\colhead{ $<W_{LSR}>$                     } &
\colhead{ $\sigma_{U}$    \tablenotemark{c}  } &
\colhead{ $\sigma_{V}$    \tablenotemark{c}  } &
\colhead{ $\sigma_{W}$    \tablenotemark{c}  } &
\colhead{ $<T>$                         } &
\colhead{ $<$[Fe/H]$>$ \tablenotemark{d}  } &
\colhead{ $\sigma$[Fe/H]\tablenotemark{e} } &
\colhead{ $\mid$Z$\mid$ \tablenotemark{f}  } \\
}

\startdata
      &     &      &    &    &      &       &   &     &    &    &      \\
 LBHB & 27&$+$10$\pm$24&$-$207$\pm$15&$-$04$\pm$18&121$\pm$16&077$\pm$10&090$\pm$12&         260$\pm$13 &$-$1.67$\pm$0.09&0.43$\pm$0.06&0.34 \\
 CBHB & 82&$-$14$\pm$14&$-$212$\pm$12&$+$13$\pm$09&121$\pm$10&106$\pm$08&078$\pm$ 6&         266$\pm$12 &$-$1.63$\pm$0.06&0.50$\pm$0.04&2.20 \\
 CA & 50&$+$28$\pm$12&$-$043$\pm$11&$-$00$\pm$07&086$\pm$09&074$\pm$07&047$\pm$ 5&         113$\pm$12 &$-$0.52$\pm$0.08&0.42$\pm$0.08&1.10 \\
      &     &      &    &    &      &       &   &     &    &    &      \\
 MWRG&81       &$\cdots$     &$\cdots$      &$\cdots$     &155$\pm$07 & 109$\pm$05   &101$\pm$05   & $\cdots$    &$-$1.92$\pm$0.05  &$\cdots$        & 0.65       \\
 HALO~2&78       &$-$17$\pm$16 &$-$187$\pm$12 &$-$05$\pm$11 &141$\pm$11 & 106$\pm$09   &094$\pm$08   & $\cdots$    &$\cdots$  &$\cdots$        & $\cdots$   \\
 THICK DISK &$\cdots$   &$\cdots$  &$-$046$\pm$05    &  $\cdots$  & 063$\pm$06 &039$\pm$04   &039$\pm$04 & $\cdots$ &$-$0.48$\pm$0.05 &0.32$\pm$0.03   & $\cdots$   \\
      &     &      &    &    &      &       &   &     &    &    &      \\
\enddata

\tablenotetext{a}{ (LBHB) Local BHB stars, (CBHB) {\it Century Survey} BHB stars
  within 3.00 kpc (CA) {\it Century Survey}
 stars of spectral type A.
 MWRG is a local sample of metal-weak red giants with [Fe/H] $<$$-$1.0 and that
 lie within 1 kpc (Kepley et al., 2007); 
 HALO~2 is a local sample of halo stars with [Fe/H]$\leq$$-$2,2 (Chiba \&
 Beers, 2000); the THICK DISK sample is from  Soubiran et al., 2003. } 
\tablenotetext{b}{Number of stars in sample. } 
\tablenotetext{c}{Dispersions of $U$, $V$ \& $W$ in km/s. } 
\tablenotetext{d}{ Mean [Fe/H] of sample.} 
\tablenotetext{e}{ Dispersion in [Fe/H] of sample.} 
\tablenotetext{f}{ Mean height of sample above Galactic plane (kpc).} 

\end{deluxetable*}

\section{ The Mean Properties of the Samples.}

Our local BHB sample (LBHB) consists of 27 stars within 1 kpc that were identified 
as BHB stars from high resolution spectra by Kinman et al. 2000 and Behr
2003; they give accurate radial velocities and [Fe/H] for these stars. 
 Their Johnson $V$ \& $B$ magnitudes were corrected for interstellar 
extinction (Schlegel et al. 1998).  Absolute magnitudes (M$_{v}$) were found 
from their $(B-V)_{0}$ color by the formula given by Preston et al. (1991) 
 and their distances calculated. Table 1 presents the mean values of their
  Galactic velocities ($<U_{LSR}>$, $<V_{LSR}>$, $<W_{LSR}>$),
 the dispersions in these velocities ($\sigma_{U}$, $\sigma_{V}$, $\sigma_{W}$),
 their mean total space velocity $<T>$, their mean metallicity $<$[Fe/H]$>$ and 
 its dispersion $\sigma$[Fe/H] and their mean height Z above the plane;
  the data for the individual stars of 
 this LBHB sample are given in Table 2 [at end of manuscript].

Two samples (CBHB \& CA) were taken from the {\it Century Survey} 
 (Brown et al. 2008)
using the classifications, distances, radial velocities and [Fe/H] given in this
paper with the {\it NOMAD} proper motions to compute the Galactic velocities of 
 these stars.
Sample CBHB consists of 82 BHB stars whose distances are less than 3.00 kpc.
 Sample CA contains 
 50 stars that were given spectral type A in the {\it Century Survey};
they are not BHB stars but presumed to be mostly stars of higher gravity.
The  mean $J_0$ magnitude of the stars in the CA sample is comparable with 
 that  of the BHB stars in sample CBHB; the errors in proper motions of the 
stars in these two  samples should therefore be similar. The mean properties of 
the two samples of 
{\it Century Survey} stars are given in Table 1 together with similar data for
halo and thick disk samples for comparison; the data for the individual stars of
the two samples of {\it Century Survey} stars are given in Table 3 [at end of manuscript].
 The distances and velocities of the CBHB and CA samples were taken from Brown et 
 al. (2008) where details are available from which their accuracies can be
 inferred.  We note that the BHB absolute magnitudes of both the LBHB and CBHB 
  samples are based on the cubic expression in $(B-V)$ given by Preston et al. 
 (1991). In the case of the LBHB sample, the $(B-V)$ were directly observed 
 colors while those used for the CBHB sample were derived from 2MASS photometry. 
  At a very rough estimate, there might be a systematic difference of as much as
 10\% between the two distance scales. The mean error of the proper motions in 
 each coordinate is $\pm$1.0 mas y$^{-1}$ for the LBHB sample, $\pm$3.1 mas 
 y$^{-1}$ for the CBHB sample and $\pm$3.4 mas y$^{-1}$ for the CA sample.

\begin{figure}
\includegraphics[width = 3.5in]{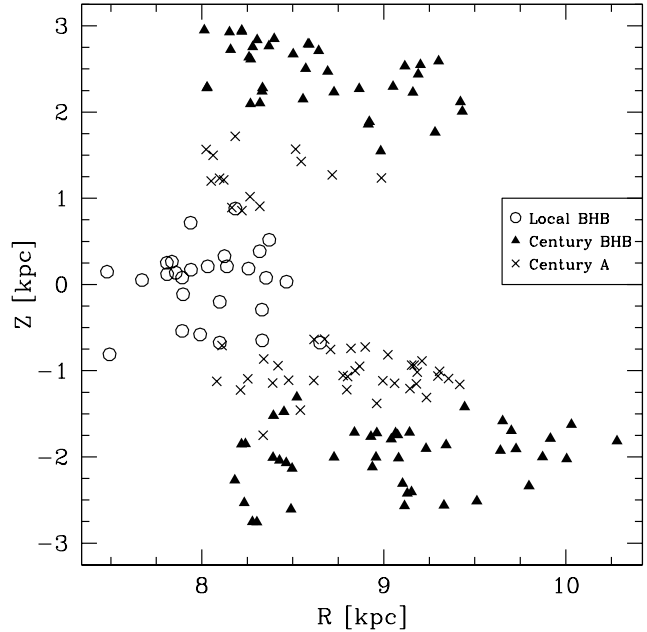}
\caption{ The Galactic spatial distributions of the local BHB sample (open circles),
 the Century BHB sample (filled triangles) and the Century A-type stars 
 (crosses). R is the galactocentric distance and Z is the height above the plane.
 We assume that the Sun is located at (R,Z) = (8,0) kpc.
\label{Fig1}}
\end{figure}

The Galactic spatial distributions of these three samples are shown in Fig. 1.
 If significant systematic 
 errors were present in the $NOMAD$ proper motions, we would expect them to 
 produce non-zero values in the mean $U$ and $W$ velocities since 
 there is no evidence that a major halo stream passes through the solar 
 neighborhood (c.f. Seabroke et al. 2008).
 The measured values of $<$U$>$ and $<$W$>$ are +10$\pm$24 and $-$04$\pm$18  
 km s$^{-1}$ for the LBHB and $-$14$\pm$14 and +13$\pm$09 km s$^{-1}$ for the
 CBHB sample respectively. We conclude that the proper motions do not contain
 a systematic error which would produce an error in $<$U$>$ and $<$W$>$ that is
 as large as 20 km s$^{-1}$. Errors in the proper motions of the CBHB sample 
 will have their greatest effect on the space velocities U and V for stars 
 at the North Galactic Pole. In this location an error of 1.0 mas y$^{-1}$ will
 produce a maximum error of 20 km s$^{-1}$ in either U or V for a star at a 
 distance of 3 kpc. We therefore consider that the likely systematic error in 
 these proper motions is not likely to produce a systematic error in V that is 
 greater than 20 km s$^{-1}$. In support of this conclusion, we note that the
 V$_{LSR}$ that we find for the LBHB and CBHB samples agree with that of the HALO2
 sample within this error. 

  We see that the $<V_{LSR}>$ of the BHB stars in 
 the LBHB and the {\it Century Survey} CBHB samples
 as well as their dispersions in $U$,$V$ \& $W$ are similar to those
 of the halo samples in Table 1. The kinematics of sample CA, on the other
 hand,  are similar to those of the thick disk sample (Soubiran et al. 2003) 
 and significantly different from those of the {\it Century Survey} BHB samples.
 The Galactic velocities of these BHB stars at $\mid$Z$\mid$ of 0.34       \&
 2.20 kpc show that, on average, they have halo rather than disk kinematics.

\subsection{ The Effects of Selection on the Samples.}

  The {\it Century Survey} stars were selected 
  photometrically and therefore have no kinematic bias. A comparison of the 
  classification of the {\it Century Survey} stars with previous classifications 
  of BHB stars at the North Galactic Pole (Kinman et al., 2008) 
  shows very good agreement and gives confidence that
  the {\it Century Survey} classifications are largely correct \footnote{ Seven
  of the BHB stars in Table 3, CHSS 833, 196, 2103, 2152, 2323, 2353 and 2372 
  have been previously classified as BHB stars; only one, CHSS 2349 (HZ 31) has 
  been classified differently (Greenstein \& Sargent, 1974; Hill et al., 1982.)}.
  The local (LBHB) sample may have some kinematic bias because many of
  its stars were selected because of their high proper motions\footnote{ 
  In Table 2, an asterisk following the star's HD or BD number shows that it was
  selected by its color alone. The remaining stars were selected by color from a 
  proper-motion limited sample.}. The bias 
  cannot be very large, however, because (to a first order)  the local and
  {\it Century Survey} BHB stars have similar kinematics (Table 1). 

  To examine the possible bias of the 27-star LBHB sample in more detail, we
  used a sample of 75 first-ascent halo red giants which were selected without
  kinematic bias. These stars all have [Fe/H]$<-1.0$, distance less than 1 
  kpc and are a subset of the local halo sample of Morrison et al. (2008).
  We randomly selected 1000 sub-samples of size 27 from this sample and 
  examined how often these had the same kinematic parameters as the LBHB
  sample. If the LBHB sample has significant kinematic selection effects, its
  parameters ($<V>, \sigma_U, \sigma_V$ and $\sigma_W$) should be different from
  those of the red giant sample. They should therefore appear only rarely in
  the 1000 sub-samples. In fact, the actual values of $<V>, \sigma_U$ and
  $\sigma_W$ appeared quite often; thus, these values appeared in 
  61\%, 8\% and 44\% of the subsamples respectively. Only the value of 
  $\sigma_V$ (77 \vkms in the LBHB sample) appeared just 1.5 \% of the time in
  the 1000 sub-samples. This shows that the LBHB sample is only significantly
  different in $\sigma_V$ (and this at a fairly low significance level) from
  the unbiased red giant sample. This difference is in the expected sense if
  the local BHB sample lacks stars with low proper motions, but it is clearly
  not a strong effect.
  The question of selection on local halo samples is discussed further in the 
  Appendix.

    We made a  similar comparison between the local red giant sample (now with
    distances less than 2.5 kpc) and the
    82-star {\it Century Survey} BHB sample. In this case, we found no 
    significant differences in the mean V velocity, $\sigma_U$. $\sigma_V$,
    and $\sigma_W$ when 1000 sub-samples of 82 stars from the red giant sample
    were compared with the BHB sample. These values or smaller ones occurred 
    8\%, 10\%, 33\% and 9\% of the time. The median $\mid$Z$\mid$ of the local
    red giant sample is 0.65 kpc and so it is closer to the plane than the 
    {\it Century Survey} BHB sample and some small difference in kinematics 
    might have been expected (Morrison et al. 2008), but none was found.

\begin{figure}
\includegraphics[width = 3.5in]{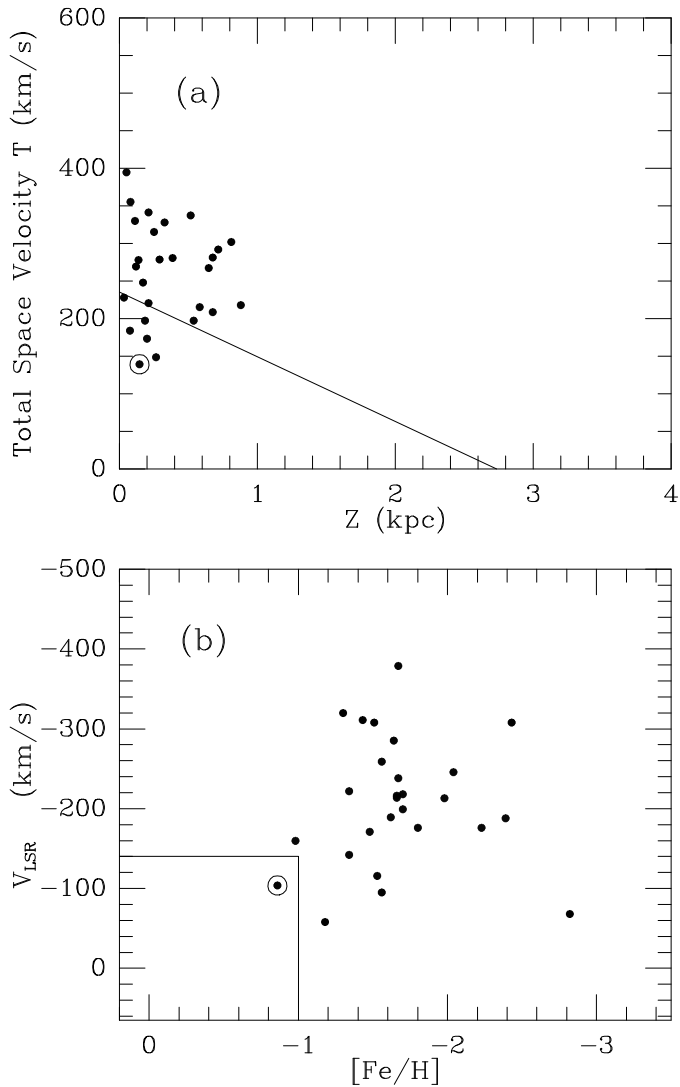}
\caption{ The local BHB Sample LBHB. The significance of the lines is described
 in Sec. 3.2. The star (HD 130201) whose Bayesian probability of belonging to the halo is less
than 0.5 (i.e. a likely disk star) is shown encircled.
\label{Fig2}}
\end{figure}

\begin{figure}
\includegraphics[width = 3.5in]{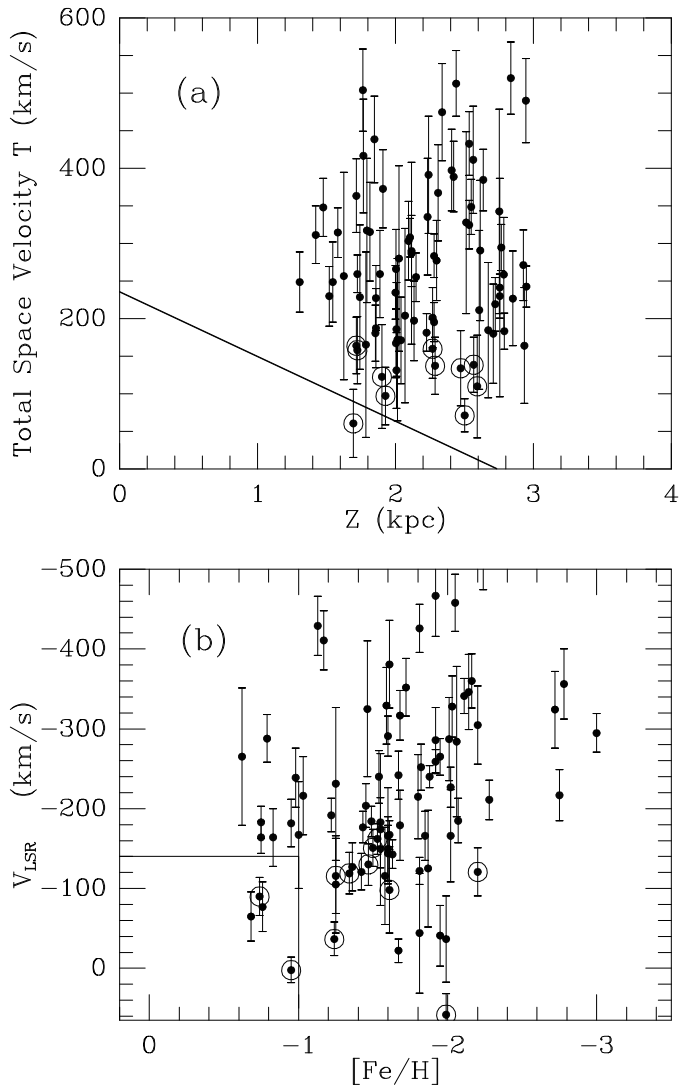}
\caption{ The {\it Century Survey} BHB sample CBHB. The significance
 of the lines is described 
 in Sec. 3.2. Stars whose Bayesian probability of belonging to the halo is less
than 0.5 (i.e. likely disk stars) are shown encircled.
\label{Fig3}}
\end{figure}

\begin{figure}
\includegraphics[width = 3.5in]{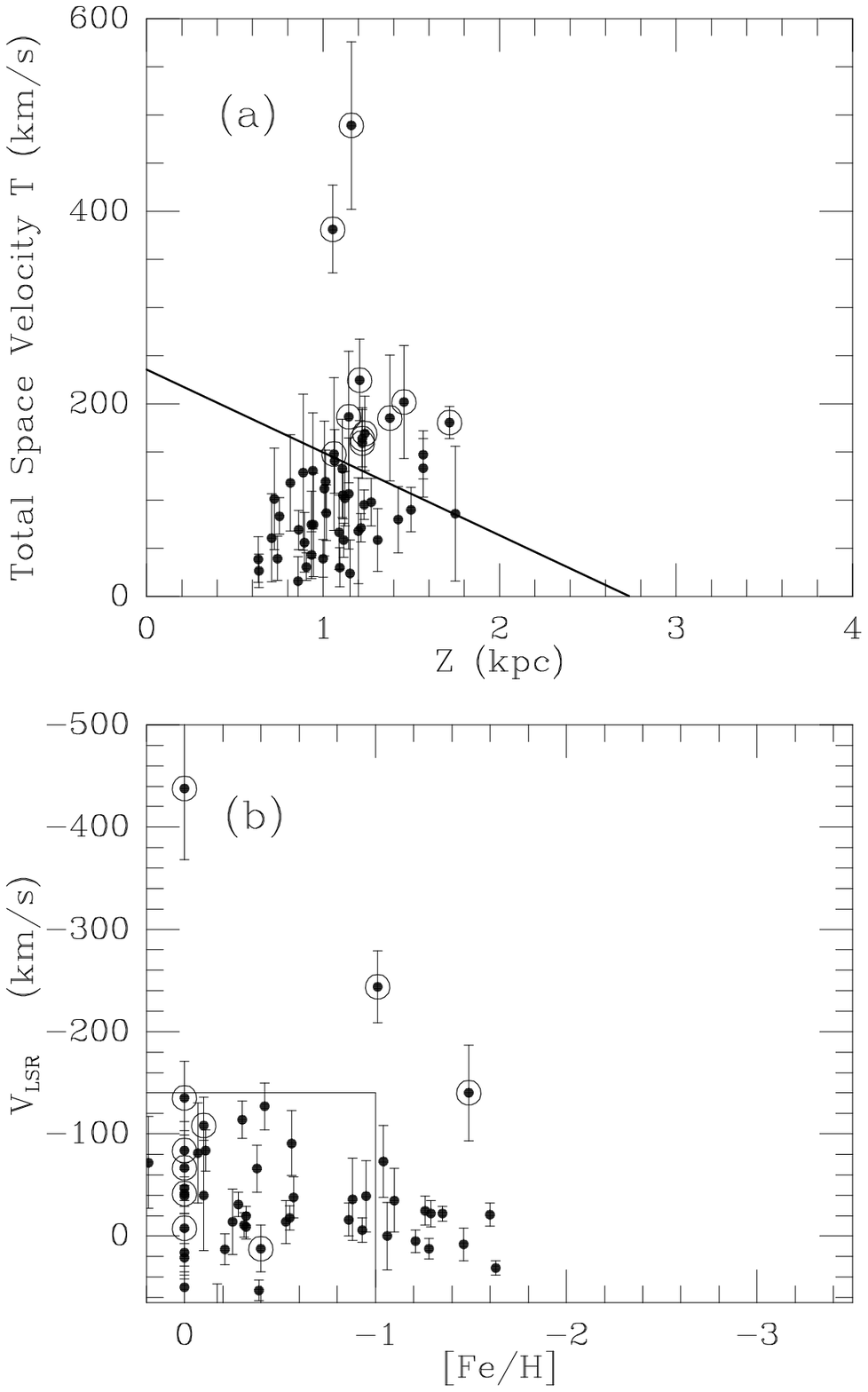}
\caption{The {\it Century Survey} A-type star sample CA. The significance 
 of the lines is described 
 in Sec. 3.2. Stars whose Bayesian probability of belonging to the halo is more 
than 0.5 (i.e. likely halo stars) are shown encircled.
\label{Fig4}}
\end{figure}

\subsection{ How many Thick Disk BHB stars could there be?}

  The earliest systematic characterizations of stellar populations (e.g. Oort,
   1958) assumed that the older populations were {\it smoothly} distributed.
  This assumption is still implicit in much work on the disk populations
  which are commonly defined by a scale heights, scale lengths and velocity 
  dispersions in orthogonal coordinates. Recent research on the halo, however,
  has shifted very largely to studies of how it departs from a smooth 
  distribution, either as overdensities or the clumping in the
  distribution of kinematic quantities that are adiabatic invariants. The latter
  approach is only possible for those relatively nearby stars that have
  well-determined distances, radial velocities and proper motions 
  (Helmi et al., 1999, Kepley et al., 2007, Morrison et al., 2008).
  For the present, when we characterize the halo by parameters 
  (such as the scale height, and velocity dispersions) that are appropriate for 
  smooth distributions, we must remember that the validity of these parameters 
  may be limited. The incompleteness of samples will also limit the conclusions
  that drawn about these parameters.
  With these {\it caveats} we use 
  three criteria for distinguishing disk from halo stars:

  (1) Martin \& Morrison (1998) showed that in a plot of the total space
  velocity (T) against $\mid$Z$\mid$ (the height above the plane), the 
  thick-disk RR Lyrae stars lie 
   below the line which has T = 235 km/s at Z = 0 kpc and T = 0 km/s at
   Z = 2.73 kpc. We give this plot for the local BHB sample
   sample, the {\it Century Survey} BHB sample and the {\it Century Survey} 
   A-type star sample in Figs 2(a), 3(a) \& 4(a) respectively.
    On this criterion, six (22\%) of the LBHB sample are
   disk stars but only one (1\%) of the CBHB sample.    
   On the other hand, the majority (76\%) of the CA sample (non-BHB A-type
   stars) belong to the disk.

  (2) Layden et al, (1996) used three different criteria to define disk stars.
    These were also used by Dambis \& Rastorguev (2001) who preferred their Disk-2
    criterion because it gave the least contamination by halo stars. We give reasons
    for agreeing with this conclusion in the Appendix (A.2). This criterion 
    defines disk stars as having [Fe/H] $\geq$ $-$1.0 and V$_{\theta}$ $>$  +80
    \vkms~ (V$_{LSR}$$\sim$$>$$-$140 \vkms). These limits are shown as the box in 
    the $V_{LSR}$ {\it vs.} [Fe/H] 
   plots for the local BHB 
   sample, the {\it Century Survey} BHB sample and the {\it Century Survey} 
   A-type star sample in Figs 2(b), 3(b) \& 4(b) respectively.
    On this criterion, only one of the LBHB sample is a thick disk member and 
    only four (5\%)
   of the {\it Century Survey} CBHB sample belongs to the thick disk, but 36 (72\%)
   of the CA sample are disk stars.
 
 (3) Venn et al. (2004) calculated the Bayesian probabilities P$_{thin}$, P$_{thick}$, P$_{halo}$
     that a star belongs to the thin disk, thick disk and halo respectively from its 
  fit to the corresponding thin disk, thick disk \& halo galactic Gaussian velocity 
 ellipsoid components
 \footnote{Thin Disk (Dehnen \& Binney,
  1998); Thick Disk (Soubiran et al.,  2003) and Halo (Chiba \& Beers,  2000).
 Similar estimations of such Bayesian probabilities have been given by 
 Mishenina et al., 2004 and Reddy et al., 2006.}. We determined 
 these Bayesian probabilities (normalized so that P$_{thin}$ + P$_{thick}$
 + P$_{halo}$ equals unity); they are  
 given in Tables 2 and 3 [at end of manuscript]. We used the same priors as Venn et al. for the thin disk and
 thick disk but used the velocity ellipsoid of the red giant sample (Morrison et al.,
 2008) for the halo. Stars that have P$_{halo}$$\leq$0.50 are shown encircled in Figs.
 2 and 3. Stars with P$_{halo}$$\geq$0.50 are shown encircled in Fig. 4.

  The mean probability P$_{disk}$ (P$_{thin}$ + P$_{thick}$) that a star belongs to 
  the disk and not the halo is 
  0.045$\pm$0.023,      
  0.133$\pm$0.028 and 0.750$\pm$0.052 for the local BHB, {\it Century Survey} BHB and
  {\it Century Survey} A-type non-BHB stars respectively. 
  Only one star (HD 130201) in the local BHB sample has both the kinematics and 
  [Fe/H] to make it likely to be a disk star.
  In the  {\it Century Survey}, there are ten BHB stars that have both
  P$_{disk}$ $>$ 0.50 but only two of these have [Fe/H]$>$$-$1.0. In the Appendix
  (A.2), we show that the use of Bayesian probabilities to select {\it disk} RR Lyrae
  stars leads to the inclusion of metal-weak stars with increasing height above the
  plane. Thus we need a metallicity restriction if we are to define the disk in
  terms of a homogeneous population. It is particularly needed for samples
  such as the {\it Century Survey} that are well outside the plane. If we take
  into account 
  the formal uncertainties of the {\it Century Survey} [Fe/H] ( $\pm$0.25 dex),
  we conclude that not more than two or three of the CBHB sample are likely to
  belong to the disk if it defined by the Disk-2 criterion of Layden et al.
  (1996). An estimate of a upper limit to the fraction disk stars in the CBHB
  sample is discussed in the Appendix.

  There are eleven A-type stars in the {\it Century Survey} CA sample whose P$_{halo}$
  exceeds 0.50. Nine of these, however, 
  have [Fe/H] $>$ $-$0.5 and so are most unlikely to belong to the halo. Also, as
  in the case of the {\it Century Survey} CBHB sample, there is a metal-weak tail to
  stars whose P$_{disk}$ exceeds 0.50. 
  Brown et al. (2008) 
  estimated the distances of their A-type stars assuming that they have the absolute 
  magnitudes of globular cluster blue stragglers of the same $(B-V)_{0}$ and [Fe/H].
  The difficulty of assigning absolute magnitudes to this probably 
  heterogeneous class of stars may well have led to less certain kinematics than
  for the {\it Century Survey} BHB stars. We presume that nearly all the stars in the
  CA sample 
  are higher-gravity A-type stars (including blue stragglers) of  the thick disk and
  the balance of the evidence agrees with this.

\section{Summary}

 We used \emph{NOMAD} proper motions for 82 BHB stars from the 
{\it Century Survey}
that are nearer than 3 kpc to get Galactic velocities $U$,$V$ \& $W$ whose 
systematic errors probably do not exceed 20 \vkms.
The mean $U$, $V$ \& $W$ and corresponding velocity dispersions of these 
{\it Century Survey} BHB stars (CBHB sample)
($<\mid$Z$\mid>$ 2.20 kpc) are very similar to those of a local (LBHB) sample 
of BHB stars ($<\mid$Z$\mid>$ 0.34 kpc) and of other local halo stars such as 
the red giant (MWRG) sample of Morrison et al. (2008). In a detailed comparison,
 the CBHB sample shows no significant difference from this MWRG  sample, but
the LBHB sample has a (99\% significant) smaller velocity dispersion 
 ($\sigma_{V}$). We show (Appendix A3) that this is probably caused by
 a lack of low proper motion stars among the more distant stars in the LBHB 
 sample.

We discuss several criteria for distinguishing between disk and halo stars. The
{\it most practical}, currently, is the Disk-2 criterion of Layden et al. 
(1966); this excludes disk membership for stars with [Fe/H] $<$ $-$1.0. Using  
 this, only two or three of the CBHB sample are likely to be belong to the disk
 and only one (HD 130201) of the LBHB sample. 
 The expected distribution of proper motions in a local $\it disk$ sample is 
 such, however, that very few disk stars would be expected in the current LBHB
 sample. If we had not put any metallicity restriction on disk membership,  
 11\% of our CBHB sample might belong to the disk. We take this to be an upper 
 limit to number of disk stars in this sample. This upper limit for the 
 percentage of disk stars in the BHB sample is much lower than the 35\% found 
 in the sample of local RR Lyrae stars (99.9\%~significance).
  A sample of 50 non-BHB stars that are classified as A-type 
 in the {\it Century Survey} ($<\mid$Z$\mid>$ 1.10 kpc), on the other hand, 
 were found to be largely disk stars. 

 The cumulative number of stars within a given distance for a population of stars that
 has the scale height (1.26 kpc) of the {\it Century Survey} BHB stars is compared 
 with those in the LBHB sample, a local RR Lyrae sample and the MWRG sample. The
 RR Lyraes within 850 to 900 pc show a similar scale height to the CBHB sample in
 agreement with previous determinations (Amrose \& Mckay, 2001; Maintz \& de Boer, 
 2005). The LBHB and MWRG samples show significant incompleteness beyond 350 and 250
pc respectively so that only limited comparisons with them are possible. All three of these
local samples show incompleteness at lower galactic latitudes when compared with the
numbers expected with the scale height given by the {\it Century Survey}. 

We therefore conclude that the CBHB stars, like the RR Lyrae stars, form a quite
flattened system (like the thick disk) near the Galactic plane 
 {\it but that most of these CBHB stars belong to the inner halo and are
not disk stars.}

\acknowledgments   
         We are very grateful to Dr Mike Irwin for providing us with his 
         program for calculating population probabilities and Dr Sabine Moehler,
         Dr Sofia Feltzing  and an anonymous referee for helpful comments.
         This research has made use of the VizieR catalogue access tool,
                 CDC, Strasbourg, France.


\clearpage

\section{APPENDIX}

\subsection{Comments on the completeness and composition of local samples.}

 Our samples consist of (a) BHB stars, (b) RR Lyrae stars and (c) metal-weak 
 red giant stars that are within 1 kpc. The BHB sample (LBHB) is that in Table 2 [at end of manuscript]
 where an asterisk after the ID shows that the star was identified as a BHB star
 by its color alone; the remainder were identified by color from among stars with proper
 motions exceeding 50 mas y$^{-1}$. The RR Lyrae sample was taken from a recent
 compilation by Maintz and de Boer (2005) of 217 RR Lyrae stars for which 
 distances and radial velocities were taken from the literature and proper 
  motions from the Hipparcos and Tycho-2 catalogues; they used these data 
 to derive the galactic orbits for these stars. We classified these RR Lyrae 
 stars as {\it disk} or {\it halo} using the Bayesian 
 probabilities described in Sec. 3.2.
 The metal-weak red giant (MWRG) sample, taken from Kepley et al. (2007), is 
 given as a halo sample in Table 1. 
 We note that four out of the ten MWRG stars  that are within 300 parsecs 
\footnote{HD 6755, HD 25532, HD 44007 \& HD 175305 (Roman, 1955)}
  were originally observed because of their high
 proper motions;  little such kinematic bias is expected in the selection of the
 more distant MWRG because they were discovered from objective prism spectra.
 Table 4 summarizes the mean properties of these samples. The {\it disk} RR Lyrae
 sample is not only significantly 
 more metal-rich than the halo RR Lyrae sample but also has a significantly different
 mean pulsation period.

\begin{figure} 
\centerline{\includegraphics[width = 2.75in]{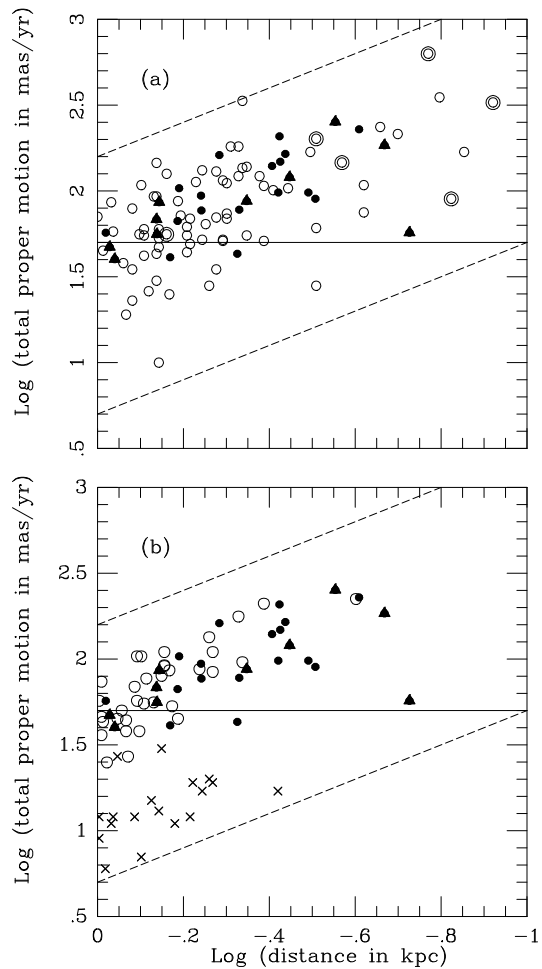}}
\caption{The ordinate is the $\log$ of the total proper motion (mas yr$^{-1}$)
 and the abscissa is the $\log$ of the distance in kpc. In both (a) and (b) 
 the local sample of BHB stars is shown by filled circles if the stars were 
 found in a proper-motion selected sample 
  and by filled triangles if they were found 
 by their colors alone.
  In (a), the metal-poor red giant sample from Kepley et al. (2007) is 
 shown by open 
 circles. A double circle indicates that the stars were identified in a proper 
 motion selected sample.
  In (b), the {\it halo} RR Lyrae stars are shown by open circles
 and the {\it disk} RR Lyrae stars by crosses. The significance of the lines is
 given in the Appendix (A.3).
\label{Fig5}}
\end{figure}

\subsection{ How well can we identify {\it disk} RR Lyrae stars?}

 Maintz and de Boer (2005) define {\it disk} RR Lyrae stars as having a 
galactic rotation ($\Theta$) greater than 100~\vkms, an orbital eccentricity less
 than 0.4 and a normalized z-extent less than 0.4 kpc. All 18 of the {\it disk} 
RR Lyrae stars that we identified by Bayesian probabilities would also be 
 identified as disk stars by these criteria except for SW Dra (which has an orbital
 eccentricity of 0.42). Among 50 RR Lyrae stars that belong to the
 disk according to the Maintz and de Boer criteria, only two (v675 Sgr and RV 
 Sex) have Bayesian probabilities that would assign them to the halo. Close to
 the galactic plane, therefore, the Bayesian probabilities seem to give an 
 adequate description of the kinematic properties. If we consider the RR Lyrae
 stars in the Maintz \& de Boer catalog that lie more than 1 kpc from the plane, 
 we find 11 stars whose Bayesian probabilities assign them to the disk.
 They have [Fe/H] in the range $-$0.45 to $-$2.23, a mean [Fe/H] of 
 $-$1.30$\pm$0.17 and a mean period of 0.538$\pm$0.024. 
 This sample of {\it disk} RR Lyrae stars ($<$$\mid$Z$\mid$$>$ = 1.37$\pm$0.11 kpc)
 is significantly different from those with $\mid$Z$\mid$ $<$1.0 (Table 4).
 This is shown in the $V$-amplitude {\it vs.} $\log$ period plot of Fig. 7 for the
 type {\it ab} RR Lyraes selected as disk members by their Bayesian probabilities.
 The curve is that for Oo I variables in the globular cluster M3 ([Fe/H] = $-$1.5)
 from Cacciari et al., (2005) who show (in their Fig. 4) that the type {\it ab}
 RR Lyrae stars 
 in halo globular clusters ([Fe/H] $<$ $-$1.0) either scatter around this line or
 lie to the right of it. RR Lyrae stars in metal-rich bulge globular clusters lie 
 well  to the right in this diagram and only the field metal-rich disk RR Lyrae stars lie
 {\it systematically} to the left of this curve (Pritzl et al., 2000). 
 The local {\it disk} RR Lyrae stars that have $\mid$Z$\mid$$<$ 1.0 kpc
  that lie to the left of the M3 curve in this diagram are all more 
 metal rich than [Fe/H] = $-$1.16 and it seems reasonable to assume that they
 comprise a more or less homogeneous population. This is not true of the 
 sample that is more distant from the plane. We conclude that we should only use
 the small sample of stars near the plane to define {\it disk} HB stars. 
 These considerations suggest that the Disk-2 definition
(Layden et al., 1996) that excludes stars more metal-weak than [Fe/H] = $-$1.0
is currently the best practical  (if somewhat conservative) definition;
 this agrees with the conclusion of Dambis \& Rastorguev (2001).

\subsection{ Proper motions as a function of distance}

In Fig. 5 we compare the proper motions as a function of distance for (a) the BHB
 and MWRG samples  and (b) the BHB and the RR Lyrae samples. The 
solid horizontal line in these $\log - \log$ plots corresponds to a proper motion
 of 50 mas y$^{-1}$ (the limit of the survey of Stetson (1991)). 
For a given velocity, the proper motion will decrease inversely with the 
distance. Consequently, in a log-log plot of the proper motion against the
distance, the points should scatter in a band of unit slope. It is seen that  
 the data lie in such bands (indicated by dashed lines) in Fig. 5.
Any systematic differences between the distance scales of our different samples 
would result in a horizontal shift of one sample with respect to another on this 
plot. Such differences appear to be small 
in the case of the {\it halo} samples.
The proper motions of the largest sample (the MWRG) in Fig. 5(a) are all 
greater than 50 mas y$^{-1}$ for stars within 0.25 kpc. The selection of BHB stars
by their proper motions will therefore miss few if any stars closer than this     
distance but could miss perhaps half the stars at a distance of 0.8 kpc. The distances
of the {\it disk} RR Lyrae stars are less certain than those of their {\it halo} 
counterparts. If we adopted the absolute magnitude M$_{v}$ = +1.11 derived from 
statistical parallaxes by Dambis \& Rastorguev (2001), the $\log$~D of these stars 
would be reduced by $\sim$0.1. Even so, a population with the kinematics of the 
{\it disk} RR Lyraes would have few stars with proper motions $>$ 50 mas y$^{-1}$ unless their
local space density is very much larger than that of the RR Lyraes.

\begin{deluxetable*}{ ccccccccccc}
\tablewidth{0cm}
\tabletypesize{\footnotesize}
\setcounter{table}{4}
\tablecaption{ Comparison of Halo Samples within 1 kpc. }
 
\tablehead{ 
\colhead{ Type \tablenotemark{a} } &
\colhead{ N \tablenotemark{b}  } &
\colhead{ N$_{high}$ \tablenotemark{c}  } &
\colhead{ N$_{low}$ \tablenotemark{d} }&
\colhead{ Range in [Fe/H]      } &
\colhead{ $<$[Fe/H]$>$                     } &
\colhead{ $<\mid$Z$\mid>$                     } &
\colhead{ $<$Ecc.$>$ \tablenotemark{e}     } &
\colhead{$<P_{ab}>$     \tablenotemark{f}  }  &
\colhead{$\rho$(d)  \tablenotemark{g}  } & 
\colhead{d    \tablenotemark{h}  } \\
}

\startdata
  RR(D)     &  18 & 7  & 11   &$-$0.07 to $-$1.34&$-$0.62$\pm$0.11&0.31$\pm$0.05 &  0.17$\pm$0.03 &0.46$\pm$0.02&6$\pm$2&0.85         \\
  RR(H)     &  34 & 20 & 14   &$-$0.71 to $-$2.43&$-$1.52$\pm$0.06&0.42$\pm$0.04 &  0.67$\pm$0.03 &0.55$\pm$0.02&12$\pm$1&0.90         \\
  LBHB      &  27 & 18 & 19   &$-$0.86 to $-$2.43&$-$1.67$\pm$0.08&0.34$\pm$0.05 &  $\cdots$      &$\cdots$ &37$\pm$15& 0.35             \\
  MWRG      &  81 & 52 & 29   &$-$1.09 to $-$3.09&$-$1.92$\pm$0.05&0.36$\pm$0.03 &  $\cdots$      &$\cdots$&190$\pm$78&0.20               \\
\enddata

\tablenotetext{a}{RR(D) Disk RR Lyrae stars; RR(H) Halo RR Lyrae stars; LBHB BHB stars
 as in Table 1; MWRG Metal-weak red giant (same as HALO1 in Table 1).}
\tablenotetext{b}{Number of stars in sample. } 
\tablenotetext{c}{Number of stars with galactic latitude $>$ 30$^{\circ}$ } 
\tablenotetext{d}{Number of stars with galactic latitude $<$ 30$^{\circ}$ } 
\tablenotetext{e}{ Mean eccentricty of Galactic orbit from Maintz \& de Boer (2007)} 
\tablenotetext{f}{ Mean period assuming the ``fundamentalized" period of type
{\it c} RR Lyraes is 1.342 times their actual period.}
\tablenotetext{g}{Local space density (stars per cubic kpc). } 
\tablenotetext{h}{distance in kpc within which $\rho$(d) was estimated. } 

\end{deluxetable*}

\begin{figure}
\centerline{\includegraphics[width = 2.75in]{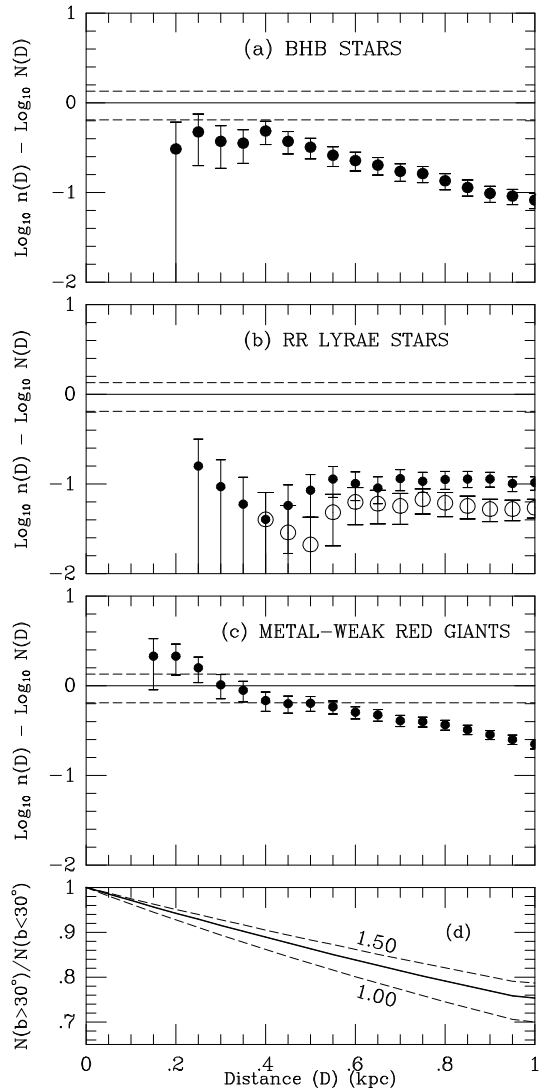}}
\caption{The ordinate is difference between $\log$ n(D) and $\log$ N(D) where
   n(D) is the number of BHB stars within a distance D  that is predicted by the
   {\it Century Survey} model and N(D) is the observed number of stars. The
   abscissa is the distance D in kpc. The plots in (a), (b) and (c) are for the
   BHB stars, RR Lyrae star and metal-weak red giants respectively. The filled
   and open circles in (b) refer to {\it halo} and {\it disk} RR Lyrae stars
   respectively. The error bars were calculated from Poisson statistics. The 
 horizontal dashed lines indicate the uncertainty in the local density that is
 predicted by the {\it Century Survey} model. Fig. 5 (d) is described in the text.
\label{Fig6}}
\end{figure}

\begin{figure}
\centerline{\includegraphics[width = 2.75in]{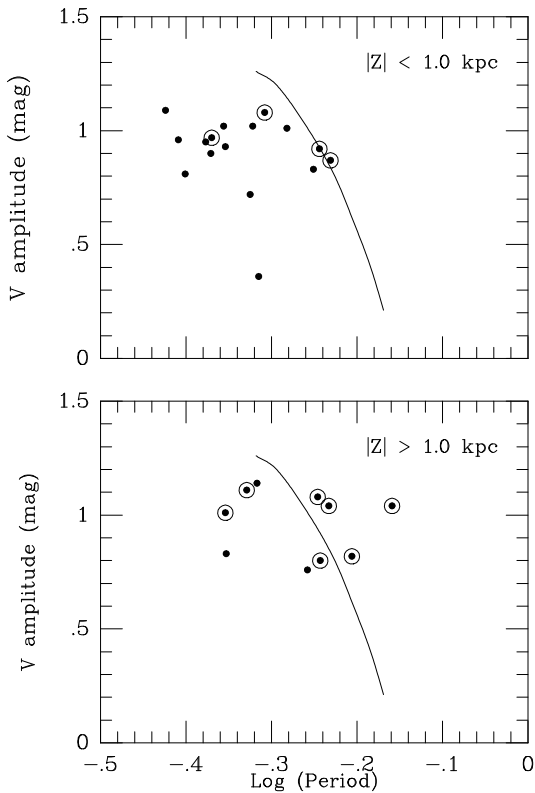}}
\caption{V-amplitude {\it vs.} $\log$(Period) for {\it disk} type-{\it ab} RR Lyrae 
 stars selected by their Bayesian probabilities for (above) stars with $\mid$ Z
$\mid$
$<$ 1 kpc and (below) stars with $\mid$Z$\mid$ $>$ 1 kpc. Stars with [Fe/H] 
 $<$ $-$1.0 
 are shown encircled. The curve is the relation for the globular cluster M3
(Cacciari et al., 2005). 
\label{Fig7}}
\end{figure}

\subsection{ A comparison with the predictions of the {\it Century Survey}.}

Brown et al. (2008) showed that the 
{\it Century Survey} BHB stars with $\mid$Z$\mid$ $<$ 4 kpc fitted an 
exponential disk with a scale height of 1.26$\pm$0.1 kpc and a local space density of 
104$\pm$37 stars kpc$^{-3}$. We computed the number (n) of BHB stars that would be 
found within a distance (D) according to this model 
and compared it with the number (N) that are actually observed.
 We show $\log$~n $-$ $\log$~N as a function of  D  in Fig. 6(a)(b)(c)
for our BHB, RR Lyrae and MWRG samples respectively. We would expect that 
 $\log$~n $-$ $\log$~N  would be constant for a given type of star if the model is
 applicable. The RR Lyrae stars show little trend of $\log$~n $-$ $\log$~N with 
distance which suggests that they have the same scale-height as the {\it Century
Survey} BHB stars. Amrose \& Mckay (2001) discussed 186 type {\it ab} RR Lyrae stars
that were found in the {\it Robotic Optical Transient Search Experiment} (ROTSE)
survey and found an exponential distribution which was consistent with scale heights
between 0.58 and 1.50 kpc. Maintz \& de Boer found a scale height that is 1.25 
to 1.30 kpc for the RR Lyraes near the plane.
  We take this as evidence that both the {\it Century 
Survey} BHB stars and RR Lyrae stars have similar scale heights. 
In the case of the BHB and MWRG samples,  $\log$~n $-$ $\log$~N is initially 
constant but then decreases with increasing distance. We take this as evidence 
that both these samples show {\it distance-dependent incompleteness}. The
alternative explanation would be that they have improbably small scale heights.
 The space
densities in Table~3 were calculated using only the range of distance over which
 $\log$~n $-$ $\log$~N  was considered constant for each sample. The errors of these
 space densities were calculated by Poisson statistics from the number of stars 
used to derive the space densities; they do not take incompleteness into account.

 We calculated the ratio of the number of
stars that would be expected to be observed at galactic latitudes (b) greater than
30$^{\circ}$ to the expected number at latitudes less than 30$^{\circ}$.
 This ratio is shown as a function of distance in Fig. 6(d) for scale heights of
 1.26 kpc (full line) and 1.00 and 1.50 kpc (dashed lines). 
 The observed numbers are given in Table 3 [at end of manuscript] and generally show fewer than 
the expected number of lower latitude stars; this is not at all unexpected since 
the crowding and increased extinction at lower latitudes will make surveys less 
effective. The problem is particularly severe for the MWRG sample because 
objective prism surveys require relatively uncrowded fields.

\begin{figure}
\centerline{\includegraphics[width = 2.75in]{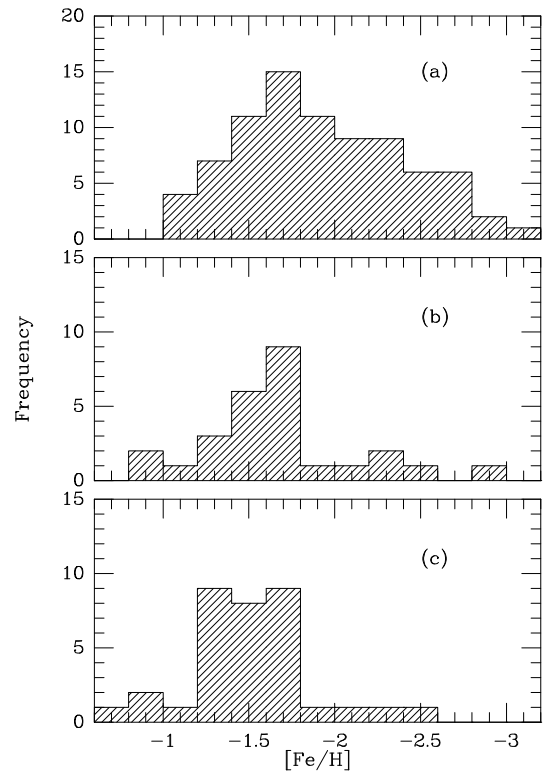}}
\caption{The distribution of [Fe/H] among different samples of halo stars within
     1 kpc. (a) MWRG sample; (b) BHB sample and (c) {\it halo} RR Lyrae sample.
\label{Fig8}}
\end{figure}

\subsection{ The distribution of [Fe/H] in the different local halo samples.}

 The distributions of [Fe/H] in the BHB and {\it halo} RR Lyrae samples are 
 sufficiently similar (Table 3 \& Fig.6) that we can combine them and the resulting
 distribution has 32 stars in the range $-$1.40$<$[Fe/H]$<$$-$1.80 and 10 that are
 more metal-poor than [Fe/H] = $-$1.80. The corresponding numbers for the MWRG 
 sample are 26 and 44. These distributions differ with a greater than 
 99.9\% significance. We conclude that either (a) the MWRG sample is deficient 
 relative to the combined HB sample in the range $-$1.40$<$[Fe/H]$<$$-$1.80, or (b) 
 the combined HB sample is very deficient in stars with [Fe/H] $<$$-$1.8, or (c) there
 are systematic differences between the metallicity scales or (d) a combination of all
 of these. A further investigation is desirable but beyond the scope of this paper.

\subsection{ The future discovery of {\it disk} BHB stars.}

 HD 130201 is the only likely disk BHB candidate. It was recognized as a BHB star by  
 Stetson (1991) from a proper-motion selected sample. It is an E-region 
 photometric standard (Menzies, et al., 1989) and, in principle could have been 
 recognized as a BHB star from its $UBV$ colors. 
  The extinction is sufficiently uncertain at this star's galactic latitude
 (+12.5$^{\circ}$), however, that a definite classification would have been 
 difficult.  Stetson, (1991) and 
 Brown et al., (2008) have pointed out that colors that would allow the 
 identification of BHB stars at high galactic latitudes or among high proper
 motion samples may give ambiguous results at lower latitudes where there are 
 large numbers Population I stars. The best hope for discovering {\it disk} BHB
 stars is therefore at higher latitudes where the identification process is 
 more effective and the expected numbers are not very much less than those at 
 lower latitudes.

\subsection{ Postscript on the Metal-Weak Thick Disk.}

 In this paper we have assumed that there are no Thick Disk stars with [Fe/H]
 $<$ $-$1.0. In other words that there is no metal-weak
 Thick Disk (MWTD). While Chiba and Beers (2000) report a 
 contribution of 30\% of the MWTD in the abundance range of $-$1.7$<$[Fe/H]$<$
 $-$1.0, \footnote{Beers et al. (2002) considered that the local fraction (within
 1 kpc) of the MWTD might be of the order of 30\% to 40\%.}
  it has been known for some time (e.g. Morrison et al. 1990) that this 
 fraction is much lower for the RR Lyrae variables.
 Ivezi\'{c} et al. (2008) find $\sim$15\% contribution for the MWTD and none more
 metal-poor than [Fe/H] = $-$1.5, but this result contains no correction for the
 accuracy of their metallicities.           
 There are eight {\it Century Survey} BHB stars ($\mid$Z$\mid$ in the range 
 1.5 to 3.0 kpc) with
 P$_{halo}$$<$0.60 and [Fe/H]$<$$-$1.50 and if all these stars are disk stars 
 then 10\% of the {\it Century  Survey} BHB sample could belong to the disk.
  To proceed further, we need 
 more precise [Fe/H] for these stars; meanwhile we suggest that 10\% be
 regarded as an {\it upper limit} to the percentage of disk stars in the {\it 
 Century Survey} BHB sample.

 A definitive discussion of the MWTD requires accurate data and a consideration
 of selection biases. Morrison et al. (2008) have discussed these biases. Their
 local sample of stars that have [Fe/H] $<$ $-$1.0 have well-defined 
 metallicities and contain almost no Thick Disk stars. Their sample, however, is 
 based on surveys at high galactic latitude, so we would expect that Thick
 Disk stars might be under-represented. 
 Reddy \& Lambert (2008) give abundance analyses for
 sixty MWTD candidates drawn from the catalogs of Ariyanto et al.
 (2005) and Schuster et al. (2006). These catalogs also contain kinematical 
 selection effects and so may under-represent the Thick Disk component.
 Reddy \& Lambert find 14 stars that might be considered
 MWTD and 20 that they call hybrid and which might be considered either disk or
 halo. {\it They were unable to identify a conclusive abundance signature that 
 would distinguish a MWTD star from a halo star}. 
 They note that the velocity characteristics of the MWTD may not be those of the
 Thick Disk. This would not be suprising if minor streams were present.
 The overall conclusion seems to be that the population of the MWTD is small
 compared with that of the Thick Disk and comparable in size to that of 
 a hybrid population of stars that cannot be conveniently classified with 
 present data.

\begin{figure}
\centerline{\includegraphics[width = 2.5in]{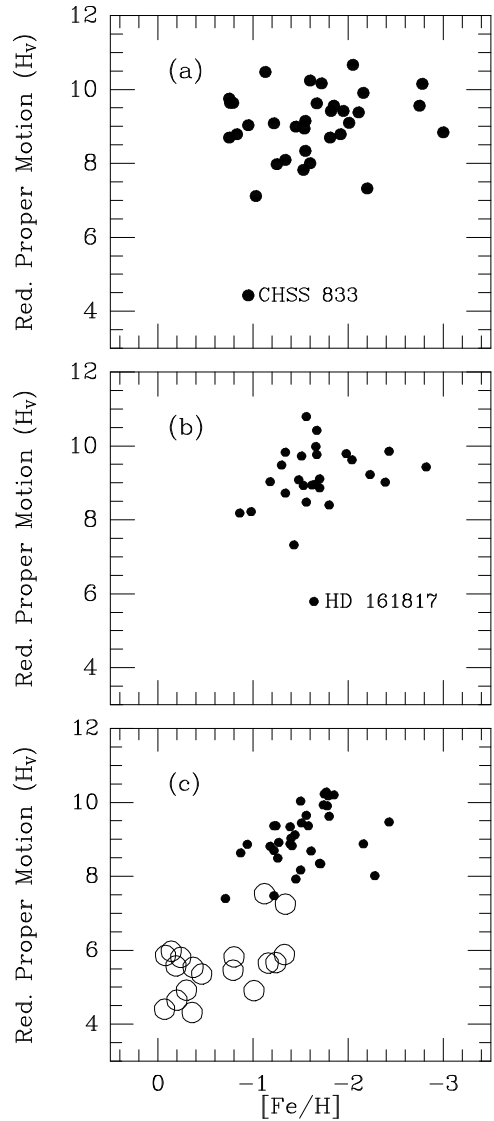}}
\caption{The reduced proper motion ($H_V$) (ordinate) {\it vs.} [Fe/H]
         (abscissa) for (a) the nearest 34 stars (D $<$ 2.75 kpc) of the
  {\it Century} CBHB sample, (b) the local BHB sample (LBHB) and (c) the
  RR Lyrae stars within one kpc. Here halo RR Lyrae stars are shown by filled
  circles and disk RR Lyrae stars by large open circles. 
\label{Fig9}}
\end{figure}

 \subsection{Postscript on the use of Reduced Proper Motions}
   The referee has asked us to consider the use of reduced proper motions for
   separating halo from thick disk stars. Stetson (1981) used this method to
   pick out BHB stars from other early type stars. More recently, Rybka (2006)
   has shown that red clump stars can be separated from intrinsically fainter 
   stars with 90\% efficiency by this method.
   Following Stetson, the reduced proper motion (H$_{V}$) is defined in terms of
   the $V$ magnitude and the total proper motion $\mu$ as follows:
   \begin{equation}  
      H_{V}  =  V +5 + 5\log\mu                             
   \end{equation}  
   If D is the distance and M$_{V}$ is the absolute magnitude, we have:
   \begin{equation}  
      V = M_{V}  -5 + 5\log D                              
   \end{equation}  
   and so:
   \begin{equation}  
      H_{V}  = M_{V} + 5\log D   + 5\log\mu               
   \end{equation}  
     For convenience, we used equation (A3) to calculate $H_{V}$ for the nearest
   34 stars of the CBHB sample and we plot this against [Fe/H] in Fig. 9(a).
   Similar plots are given in Fig. 9 (b) for the Local BHB sample and in Fig.
   9 (c) for the Local RR Lyrae stars (where the disk RR Lyrae stars are 
   shown by large open circles). We see that the reduced proper motion ($H_{V}$)
   affords a partial separation of the halo and disk RR Lyrae stars. HD 161817 
   would be considered a disk star because it has a small proper motion but
   a high radial velocity. Only one of CBHB sample (CHSS 833) has the $H_{V}$ of
   a disk star which supports our conclusion that only a few percent of the
   {\it Century} BHB sample belongs to the disk.  


\clearpage

\begin{deluxetable*}{ ccccccccccc }
\tablewidth{0cm}
\tabletypesize{\footnotesize}
\setcounter{table}{2}
\tablecaption{ Data for BHB stars within 1 kpc (Sample LBHB) }
 
\tablehead{ 
\colhead{  HD/BD } &
\colhead{ $U_{LSR}$                     } &
\colhead{ $V_{LSR}$                     } &
\colhead{ $W_{LSR}$                     } &
\colhead{ T \tablenotemark{a}  } & 
\colhead{ $$[Fe/H]$$                    } &
\colhead{ $\mid$Z$\mid$ \tablenotemark{b}  } & 
\colhead{ D \tablenotemark{c}  } & 
\colhead{ P$_{thin}$\tablenotemark{d}  } &
\colhead{ P$_{thick}$\tablenotemark{e}  } &
\colhead{ P$_{halo}$\tablenotemark{f}  }   \\
}

\startdata
 2857$^{\ast}$  & 164$\pm$14&$-$218$\pm$17&  68$\pm$10&281$\pm$24 & $-$1.70&  0.68& 0.73&0.00$\pm$ 0.00&0.00$\pm$ 0.00&1.00$\pm$ 0.00  \\
 4850  &$-$186$\pm$19&$-$58$\pm$8&  31$\pm$5 &197 $\pm$21& $-$1.18&  0.54& 0.57&0.00$\pm$ 0.00&0.03$\pm$ 0.05&0.97$\pm$ 0.05  \\
 8376 & $-$261$\pm$9&$-$68$\pm$17&$-$69$\pm$3&278$\pm$26 & $-$2.82&  0.29& 0.57&0.00$\pm$ 0.00&0.00$\pm$ 0.00&1.00$\pm$ 0.00  \\
13780 &  176$\pm$17& $-$116$\pm$11&  45$\pm$8&216$\pm$22 & $-$1.53&  0.58& 0.65&0.00$\pm$ 0.00&0.01$\pm$ 0.01&0.99$\pm$ 0.01  \\
14829$^{\ast}$ &  111$\pm$6& $-$188$\pm$20& 154$\pm$5&267$\pm$21  &$-$2.39&   0.65& 0.73&0.00$\pm$ 0.00&0.00$\pm$0.00 &1.00$\pm$ 0.00  \\
31943 &   55$\pm$7& $-$160$\pm$11&  37$\pm$9&173$\pm$16  & $-$0.98&  0.20& 0.32&0.00$\pm$ 0.00&0.31$\pm$ 0.04&0.69$\pm$ 0.04  \\
252940 &$-$131$\pm$5&$-$176$\pm$17& 63$\pm$6&228$\pm$19  & $-$1.80&  0.03& 0.47&0.00$\pm$ 0.00&0.00$\pm$ 0.00&1.00$\pm$ 0.00  \\
60778$^{\ast}$  &  47$\pm$8& $-$142$\pm$13&$-$107$\pm$12&184$\pm$19 &$-$1.34& 0.08& 0.45&0.00$\pm$ 0.00&0.04$\pm$ 0.05&0.96$\pm$ 0.05  \\
74721$^{\ast}$  &   7$\pm$4& $-$171$\pm$16&$-$98$\pm$12&197$\pm$20   & $-$1.48& 0.19& 0.36&0.00$\pm$ 0.00&0.03$\pm$ 0.05&0.97$\pm$ 0.05  \\
78913 &  107$\pm$3&$-$311$\pm$5& 23$\pm$9 &330$\pm$11   & $-$1.43&     0.11& 0.47&0.00$\pm$ 0.00&0.00$\pm$ 0.00&1.00$\pm$ 0.00 \\
86986$^{\ast}$ &  259$\pm$25&$-$216$\pm$22& 52$\pm$5&341$\pm$34& $-$1.66&    0.21& 0.28&0.00$\pm$ 0.00&0.00$\pm$ 0.00&1.00$\pm$ 0.00 \\
87047 &  38$\pm$11&$-$308$\pm$29& 132$\pm$4 &337$\pm$31  &  $-$2.43&    0.52& 0.65&0.00$\pm$ 0.00&0.00$\pm$ 0.00&1.00$\pm$ 0.00 \\
87112 &   94$\pm$5&$-$259$\pm$22&$-$53$\pm$8&281$\pm$24  & $-$1.56&   0.38& 0.52&0.00$\pm$ 0.00&0.00$\pm$ 0.00&1.00$\pm$ 0.00 \\
93329 &  2$\pm$6&$-$320$\pm$23&  73$\pm$11&328$\pm$26 & $-$1.30&     0.33& 0.39&0.00$\pm$ 0.00&0.00$\pm$ 0.00&1.00$\pm$ 0.00 \\
106304 & $-$22$\pm$8&$-$222$\pm$14&$-$166$\pm$22&278$\pm$27  &$-$1.34&0.14& 0.38&0.00$\pm$ 0.00&0.00$\pm$ 0.00&1.00$\pm$ 0.00 \\
+42~2309$^{\ast}$ & 28$\pm$5&$-$189$\pm$18&$-$105$\pm$6&218$\pm$20   &$-$1.62&  0.88& 0.91&0.00$\pm$ 0.00&0.00$\pm$ 0.01&1.00$\pm$ 0.01 \\
109995$^{\ast}$  &  4$\pm$3& $-$199$\pm$18&$-$95$\pm$5&221$\pm$19  &$-$1.70&   0.21& 0.22&0.00$\pm$ 0.00&0.01$\pm$ 0.01&0.99$\pm$ 0.01 \\
+25~2602$^{\ast}$& $-$195$\pm$21&$-$213$\pm$22&$-$43$\pm$5&292$\pm$31 &$-$1.98&0.72&0.72&0.00$\pm$ 0.00&0.00$\pm$ 0.00&1.00$\pm$ 0.00 \\
117880  & 52$\pm$5&$-$308$\pm$24& $-$43$\pm$15&315$\pm$29 & $-$1.51& 0.25& 0.37&0.00$\pm$ 0.00&0.00$\pm$ 0.00&1.00$\pm$ 0.00 \\
128801  & 41$\pm$8&$-$95$\pm$10& $-$107$\pm$6&149$\pm$14  & $-$1.56&  0.26& 0.31&0.00$\pm$ 0.00&0.23$\pm$ 0.10&0.77$\pm$ 0.10 \\
130095 & $-$79$\pm$15&$-$246$\pm$22& 75$\pm$4&269$\pm$27 & $-$2.04&  0.12& 0.25&0.00$\pm$ 0.00&0.00$\pm$ 0.00&1.00$\pm$ 0.00 \\
130201 &  43$\pm$5&$-$104$\pm$8&$-$82$\pm$11&139$\pm$14 &  $-$0.86&   0.15& 0.68&0.00$\pm$ 0.00&0.52$\pm$ 0.20&0.48$\pm$ 0.20 \\
139961 & $-$44$\pm$18&$-$379$\pm$32& 100$\pm$8&394$\pm$38 & $-$1.67& 0.05& 0.38&0.00$\pm$ 0.00&0.00$\pm$ 0.00&1.00$\pm$ 0.00 \\
161817$^{\ast}$& $-$168$\pm$4&$-$285$\pm$5&$-$129$\pm$3&355$\pm$7   & $-$1.64& 0.08& 0.19&0.00$\pm$ 0.00&0.00$\pm$ 0.00&1.00$\pm$ 0.00 \\
167105& 125$\pm$15&$-$214$\pm$8&  2$\pm$8&248$\pm$19 & $-$1.66&      0.17& 0.38&0.00$\pm$ 0.00&0.00$\pm$ 0.00&1.00$\pm$ 0.00 \\
213468 & $-$94$\pm$6&$-$238$\pm$26& 161$\pm$5&302$\pm$27  & $-$1.67&  0.81& 0.96&0.00$\pm$ 0.00&0.00$\pm$ 0.00&1.00$\pm$ 0.00 \\
+01~0548$^{\ast}$ & 110$\pm$7&$-$176$\pm$19&$-$22$\pm$8&208$\pm$22 & $-$2.23&   0.68& 0.94&0.00$\pm$ 0.00&0.04$\pm$ 0.03&0.96$\pm$ 0.03 \\
\enddata

\tablenotetext{a}{  Total Space Velocity (km/s).} 
\tablenotetext{b}{  Height of star above Galactic plane (kpc).} 
\tablenotetext{c}{ Distance of star (kpc).} 
\tablenotetext{d}{ Probability that star belongs to thin disk. } 
\tablenotetext{e}{ Probability that star belongs to thick disk. } 
\tablenotetext{f}{ Probability that star belongs to the halo. } 

\end{deluxetable*}

\clearpage
\LongTables

\begin{deluxetable}{ cccccccccccc}
\tablewidth{0cm}
\tabletypesize{\footnotesize}
\setcounter{table}{3}
\tablecaption{ Data for {\it Century Survey} Stars (Samples CBHB \& CA.) }
 
\tablehead{ 
\colhead{  CHSS  } &
\colhead{ $U_{LSR}$                     } &
\colhead{ $V_{LSR}$                     } &
\colhead{ $W_{LSR}$                     } &
\colhead{ T \tablenotemark{a}  } & 
\colhead{ $$[Fe/H]$$                    } &
\colhead{ $\mid$Z$\mid$ \tablenotemark{b}  } & 
\colhead{ D \tablenotemark{c}  } & 
\colhead{ P$_{thin}$\tablenotemark{d}  } &
\colhead{ P$_{thick}$\tablenotemark{e}  } &
\colhead{ P$_{halo}$\tablenotemark{f}  } & 
\colhead{ Class }   \\
}

\startdata
Sample CBHB &      &    &    &      &       &   &     &    &    &     &    \\
 3107  & $-$028$\pm$37 & $-$164$\pm$36 & $+$040$\pm$26 & 171$\pm$58 & $-$0.83& 2.04 & 2.52 &  0.00$\pm$0.00 & 0.35$\pm$0.25& 0.65$\pm$0.25&BHB\\
 3110  & $+$073$\pm$36 & $-$183$\pm$31 & $+$005$\pm$24 & 197$\pm$53 & $-$1.55& 2.13 & 2.65 &  0.00$\pm$0.00 & 0.11$\pm$0.04& 0.89$\pm$0.04&BHB\\
 3034  & $+$163$\pm$25 & $-$183$\pm$20 & $-$043$\pm$24 & 249$\pm$40 & $-$0.75& 1.31 & 1.86 &  0.00$\pm$0.00 & 0.00$\pm$0.00& 1.00$\pm$0.00&BHB\\
 3048  & $-$131$\pm$28 & $-$458$\pm$36 & $-$164$\pm$30 & 504$\pm$55 & $-$2.05& 1.76 & 2.62 &  0.00$\pm$0.00 & 0.00$\pm$0.00& 1.00$\pm$0.00&BHB\\
 3052  & $+$074$\pm$63 & $-$216$\pm$49 & $+$013$\pm$53 & 229$\pm$96 & $-$1.03& 1.74 & 2.75 &  0.00$\pm$0.00 & 0.01$\pm$0.00& 0.99$\pm$0.00&BHB\\
 3276  & $-$082$\pm$64 & $-$287$\pm$52 & $+$107$\pm$49 & 317$\pm$96 & $-$2.01& 1.79 & 2.66 &  0.00$\pm$0.00 & 0.00$\pm$0.00& 1.00$\pm$0.00&BHB\\
 3218  & $-$151$\pm$32 & $-$242$\pm$30 & $-$053$\pm$18 & 290$\pm$47 & $-$1.67& 2.12 & 2.63 &  0.00$\pm$0.00 & 0.00$\pm$0.00& 1.00$\pm$0.00&BHB\\
 3335  & $-$046$\pm$27 & $-$360$\pm$34 & $-$023$\pm$23 & 364$\pm$49 & $-$2.16& 1.72 & 2.42 &  0.00$\pm$0.00 & 0.00$\pm$0.00& 1.00$\pm$0.00&BHB\\
 3411  & $+$046$\pm$17 & $-$295$\pm$24 & $+$099$\pm$15 & 315$\pm$33 & $-$3.00& 1.58 & 2.57 &  0.00$\pm$0.00 & 0.00$\pm$0.00& 1.00$\pm$0.00&BHB\\
 3877  & $-$335$\pm$45 & $-$150$\pm$40 & $+$016$\pm$22 & 367$\pm$64 & $-$1.60& 2.31 & 2.59 &  0.00$\pm$0.00 & 0.00$\pm$0.00& 1.00$\pm$0.00&BHB\\
 3886  & $+$309$\pm$29 & $-$217$\pm$32 & $+$093$\pm$19 & 387$\pm$47 & $-$2.75& 2.42 & 2.68 &  0.00$\pm$0.00 & 0.00$\pm$0.00& 1.00$\pm$0.00&BHB\\
 3689  & $+$022$\pm$42 & $-$116$\pm$47 & $-$034$\pm$28 & 123$\pm$69 & $-$1.25& 1.90 & 2.32 &  0.00$\pm$0.04 & 0.87$\pm$0.34& 0.13$\pm$0.38&BHB\\
 3888  & $-$169$\pm$28 & $-$356$\pm$44 & $-$053$\pm$16 & 398$\pm$55 & $-$2.78& 2.41 & 2.67 &  0.00$\pm$0.00 & 0.00$\pm$0.00& 1.00$\pm$0.00&BHB\\
 3528  & $+$116$\pm$15 & $-$288$\pm$30 & $+$027$\pm$19 & 312$\pm$39 & $-$0.79& 1.42 & 2.11 &  0.00$\pm$0.00 & 0.00$\pm$0.00& 1.00$\pm$0.00&BHB\\
 1676  & $-$116$\pm$17 & $-$204$\pm$27 & $-$006$\pm$16 & 235$\pm$36 & $-$1.45& 2.00 & 2.74 &  0.00$\pm$0.00 & 0.01$\pm$0.03& 0.99$\pm$0.03&BHB\\
 4003  & $+$152$\pm$65 & $-$174$\pm$95 & $+$112$\pm$76 & 257$\pm$138 & $-$1.55& 1.62 & 2.67 &  0.00$\pm$0.00 & 0.00$\pm$0.00& 1.00$\pm$0.00&BHB\\
 1786  & $+$065$\pm$23 & $-$182$\pm$30 & $+$056$\pm$13 & 201$\pm$40 & $-$0.95& 2.27 & 2.50 &  0.00$\pm$0.00 & 0.05$\pm$0.02& 0.95$\pm$0.02&BHB\\
 1851  & $-$036$\pm$36 & $-$121$\pm$30 & $+$045$\pm$18 & 134$\pm$50 & $-$2.20& 2.47 & 2.73 &  0.00$\pm$0.00 & 0.77$\pm$0.26& 0.23$\pm$0.26&BHB\\
 1930  & $+$026$\pm$09 & $-$252$\pm$29 & $+$030$\pm$11 & 255$\pm$32 & $-$1.82& 2.15 & 2.27 &  0.00$\pm$0.00 & 0.00$\pm$0.00& 1.00$\pm$0.00&BHB\\
 0833  & $+$049$\pm$11 & $+$002$\pm$16 & $-$052$\pm$10 & 071$\pm$22 & $-$0.95& 2.50 & 2.57 &  0.33$\pm$0.21 & 0.65$\pm$0.19& 0.02$\pm$0.01&BHB\\
 2181  & $-$222$\pm$40 & $-$166$\pm$31 & $-$006$\pm$17 & 277$\pm$53 & $-$1.85& 2.30 & 2.72 &  0.00$\pm$0.00 & 0.00$\pm$0.00& 1.00$\pm$0.00&BHB\\
 2183  & $-$114$\pm$15 & $-$352$\pm$36 & $+$104$\pm$12 & 384$\pm$41 & $-$1.72& 2.64 & 2.68 &  0.00$\pm$0.00 & 0.00$\pm$0.00& 1.00$\pm$0.00&BHB\\
 2233  & $+$063$\pm$25 & $-$119$\pm$26 & $-$026$\pm$12 & 137$\pm$38 & $-$1.34& 2.29 & 2.36 &  0.00$\pm$0.00 & 0.77$\pm$0.13& 0.23$\pm$0.13&BHB\\
 2305  & $-$129$\pm$17 & $-$122$\pm$17 & $+$082$\pm$10 & 196$\pm$26 & $-$1.81& 2.28 & 2.30 &  0.00$\pm$0.00 & 0.02$\pm$0.02& 0.98$\pm$0.02&BHB\\
 2366  & $+$288$\pm$48 & $-$286$\pm$41 & $-$093$\pm$42 & 416$\pm$76 & $-$1.92& 1.77 & 2.70 &  0.00$\pm$0.00 & 0.00$\pm$0.00& 1.00$\pm$0.00&BHB\\
 2522  & $-$222$\pm$42 & $-$077$\pm$31 & $+$081$\pm$10 & 249$\pm$53 & $-$0.76& 1.55 & 2.32 &  0.00$\pm$0.00 & 0.00$\pm$0.00& 1.00$\pm$0.00&BHB\\
 2523  & $+$099$\pm$11 & $-$265$\pm$23 & $+$018$\pm$14 & 283$\pm$29 & $-$1.95& 2.28 & 2.48 &  0.00$\pm$0.00 & 0.00$\pm$0.00& 1.00$\pm$0.00&BHB\\
 2607  & $+$094$\pm$36 & $-$240$\pm$33 & $+$028$\pm$31 & 259$\pm$58 & $-$1.54& 1.89 & 2.63 &  0.00$\pm$0.00 & 0.00$\pm$0.00& 1.00$\pm$0.00&BHB\\
 2934  & $-$070$\pm$21 & $-$341$\pm$22 & $-$008$\pm$24 & 348$\pm$39 & $-$2.11& 1.48 & 2.42 &  0.00$\pm$0.00 & 0.00$\pm$0.00& 1.00$\pm$0.00&BHB\\
 3075  & $+$024$\pm$26 & $-$164$\pm$20 & $-$072$\pm$18 & 181$\pm$37 & $-$0.75& 1.85 & 2.46 &  0.00$\pm$0.00 & 0.14$\pm$0.17& 0.86$\pm$0.17&BHB\\
 3083  & $+$072$\pm$32 & $-$429$\pm$37 & $-$054$\pm$30 & 438$\pm$57 & $-$1.13& 1.85 & 2.37 &  0.00$\pm$0.00 & 0.00$\pm$0.00& 1.00$\pm$0.00&BHB\\
 3096  & $-$095$\pm$29 & $-$192$\pm$21 & $+$083$\pm$17 & 230$\pm$40 & $-$1.22& 1.52 & 2.09 &  0.00$\pm$0.00 & 0.00$\pm$0.00& 1.00$\pm$0.00&BHB\\
 3100  & $-$116$\pm$44 & $-$144$\pm$35 & $-$017$\pm$27 & 186$\pm$62 & $-$1.60& 2.01 & 2.53 &  0.00$\pm$0.00 & 0.14$\pm$0.40& 0.86$\pm$0.40&BHB\\
 2996  & $+$023$\pm$24 & $-$162$\pm$19 & $-$014$\pm$22 & 164$\pm$38 & $-$1.53& 1.72 & 2.73 &  0.00$\pm$0.00 & 0.51$\pm$0.22& 0.49$\pm$0.22&BHB\\
 3019  & $-$017$\pm$14 & $-$259$\pm$15 & $+$001$\pm$14 & 260$\pm$25 & $-$1.92& 1.72 & 2.93 &  0.00$\pm$0.00 & 0.00$\pm$0.00& 1.00$\pm$0.00&BHB\\
 3029  & $+$153$\pm$34 & $-$037$\pm$21 & $-$013$\pm$20 & 158$\pm$45 & $-$1.24& 1.72 & 2.76 &  0.20$\pm$0.02 & 0.31$\pm$0.28& 0.49$\pm$0.27&BHB\\
 3041  & $+$130$\pm$33 & $-$185$\pm$28 & $-$140$\pm$30 & 266$\pm$53 & $-$2.07& 2.01 & 2.92 &  0.00$\pm$0.00 & 0.00$\pm$0.00& 1.00$\pm$0.00&BHB\\
 3799  & $+$059$\pm$101& $-$231$\pm$96 & $+$037$\pm$41 & 241$\pm$145& $-$1.25& 2.76 & 2.98 &  0.00$\pm$0.00 & 0.00$\pm$0.00& 1.00$\pm$0.00&BHB\\
 1626  & $+$126$\pm$74 & $-$167$\pm$67 & $+$032$\pm$35 & 212$\pm$106& $-$1.00& 2.61 & 2.94 &  0.00$\pm$0.00 & 0.02$\pm$0.00& 0.98$\pm$0.00&BHB\\
 3803  & $-$022$\pm$91 & $-$284$\pm$94 & $-$190$\pm$37 & 342$\pm$136& $-$2.06& 2.75 & 2.93 &  0.00$\pm$0.00 & 0.00$\pm$0.00& 1.00$\pm$0.00&BHB\\
 3294  & $+$135$\pm$72 & $-$105$\pm$61 & $-$018$\pm$53 & 172$\pm$108& $-$1.25& 2.01 & 2.76 &  0.00$\pm$0.00 & 0.24$\pm$0.03& 0.76$\pm$0.03&BHB\\
 3299  & $-$110$\pm$32 & $-$127$\pm$30 & $-$083$\pm$29 & 187$\pm$53 & $-$1.36& 1.86 & 2.90 &  0.00$\pm$0.00 & 0.05$\pm$0.32& 0.95$\pm$0.32&BHB\\
 1641  & $-$010$\pm$23 & $-$130$\pm$26 & $+$047$\pm$12 & 139$\pm$37 & $-$1.47& 2.57 & 2.90 &  0.00$\pm$0.00 & 0.73$\pm$0.19& 0.27$\pm$0.19&BHB\\
 3383  & $-$017$\pm$22 & $-$090$\pm$24 & $-$032$\pm$20 & 097$\pm$38 & $-$0.74& 1.93 & 2.84 &  0.00$\pm$0.05 & 0.94$\pm$0.07& 0.06$\pm$0.12&BHB\\
 3395 & $-$169$\pm$28 & $-$328$\pm$38 & $-$052$\pm$22 & 373$\pm$52 & $-$2.03& 1.91 & 2.86 &  0.00$\pm$0.00 & 0.00$\pm$0.00& 1.00$\pm$0.00&BHB\\
 3416  & $+$148$\pm$69 & $-$044$\pm$75 & $-$059$\pm$69 & 165$\pm$123& $-$1.81& 1.78 & 2.93 &  0.00$\pm$0.00 & 0.19$\pm$0.00& 0.81$\pm$0.00&BHB\\
 3880  & $-$129$\pm$39 & $-$381$\pm$55 & $-$085$\pm$23 & 411$\pm$071& $-$1.61& 2.56 & 2.93 &  0.00$\pm$0.00 & 0.00$\pm$0.00& 1.00$\pm$0.00&BHB\\
 1652  & $-$042$\pm$73 & $-$325$\pm$85 & $+$002$\pm$45 & 328$\pm$121& $-$1.46& 2.51 & 2.99 &  0.00$\pm$0.00 & 0.00$\pm$0.00& 1.00$\pm$0.00&BHB\\
 3535  & $+$066$\pm$62 & $-$265$\pm$86 & $-$061$\pm$63 & 280$\pm$123& $-$0.62& 2.02 & 2.92 &  0.00$\pm$0.00 & 0.00$\pm$0.01& 1.00$\pm$0.01&BHB\\
 3927  & $-$320$\pm$34 & $-$346$\pm$47 & $-$056$\pm$28 & 475$\pm$064& $-$2.14& 2.34 & 2.95 &  0.00$\pm$0.00 & 0.00$\pm$0.00& 1.00$\pm$0.00&BHB\\
 3638  & $+$066$\pm$28 & $-$305$\pm$49 & $+$047$\pm$34 & 316$\pm$066& $-$2.20& 1.82 & 2.92 &  0.00$\pm$0.00 & 0.00$\pm$0.00& 1.00$\pm$0.00&BHB\\
 1699  & $+$014$\pm$26 & $+$058$\pm$26 & $-$011$\pm$26 & 061$\pm$045& $-$1.99& 1.70 & 2.90 &  0.60$\pm$0.20 & 0.39$\pm$0.19& 0.01$\pm$0.01&BHB\\
 1717  & $-$007$\pm$15 & $-$512$\pm$38 & $-$029$\pm$15 & 513$\pm$044& $-$2.24& 2.44 & 2.80 &  0.00$\pm$0.00 & 0.00$\pm$0.00& 1.00$\pm$0.00&BHB\\
 0045  & $+$045$\pm$37 & $-$098$\pm$54 & $+$020$\pm$20 & 110$\pm$068& $-$1.61& 2.59 & 2.93 &  0.00$\pm$0.07 & 0.92$\pm$0.07& 0.08$\pm$0.14&BHB\\
 0786  & $+$141$\pm$17 & $-$317$\pm$31 & $+$034$\pm$10 & 349$\pm$037& $-$1.68& 2.55 & 2.86 &  0.00$\pm$0.00 & 0.00$\pm$0.00& 1.00$\pm$0.00&BHB\\
 1830  & $-$214$\pm$17 & $-$227$\pm$25 & $+$091$\pm$12 & 325$\pm$033& $-$2.02& 2.53 & 2.77 &  0.00$\pm$0.00 & 0.00$\pm$0.00& 1.00$\pm$0.00&BHB\\
 1955  & $+$015$\pm$30 & $-$065$\pm$31 & $+$113$\pm$26 & 131$\pm$050& $-$0.68& 2.01 & 2.81 &  0.00$\pm$0.00 & 0.37$\pm$0.25& 0.63$\pm$0.25&BHB\\
 1982  & $-$136$\pm$62 & $-$116$\pm$61 & $-$046$\pm$23 & 185$\pm$090& $-$1.58& 2.67 & 2.83 &  0.00$\pm$0.08 & 0.09$\pm$0.40& 0.91$\pm$0.47&BHB\\
 2078  & $-$235$\pm$77 & $-$125$\pm$73 & $+$107$\pm$58 & 287$\pm$121& $-$1.87& 2.12 & 2.92 &  0.00$\pm$0.00 & 0.00$\pm$0.00& 1.00$\pm$0.00&BHB\\
 0196  & $-$067$\pm$12 & $-$143$\pm$18 & $+$093$\pm$10 & 183$\pm$024& $-$1.63& 2.79 & 2.86 &  0.00$\pm$0.00 & 0.06$\pm$0.02& 0.94$\pm$0.02&BHB\\
 2103  & $-$139$\pm$52 & $-$215$\pm$53 & $+$039$\pm$15 & 259$\pm$076& $-$1.80& 2.79 & 2.85 &  0.00$\pm$0.00 & 0.00$\pm$0.04& 1.00$\pm$0.04&BHB\\
 2122  & $-$006$\pm$22 & $-$211$\pm$25 & $+$059$\pm$13 & 219$\pm$036& $-$2.28& 2.72 & 2.90 &  0.00$\pm$0.00 & 0.02$\pm$0.02& 0.98$\pm$0.02&BHB\\
 2152  & $-$132$\pm$43 & $-$179$\pm$44 & $+$045$\pm$13 & 227$\pm$063& $-$1.68& 2.85 & 2.90 &  0.00$\pm$0.00 & 0.01$\pm$0.10& 0.99$\pm$0.10&BHB\\
 2161  & $+$043$\pm$28 & $-$411$\pm$37 & $+$316$\pm$12 & 520$\pm$048& $-$1.17& 2.84 & 2.90 &  0.00$\pm$0.00 & 0.00$\pm$0.00& 1.00$\pm$0.00&BHB\\
 2323  & $-$124$\pm$27 & $-$239$\pm$37 & $-$031$\pm$10 & 271$\pm$047& $-$0.98& 2.93 & 2.93 &  0.00$\pm$0.00 & 0.00$\pm$0.01& 1.00$\pm$0.01&BHB\\
 2349  & $-$114$\pm$21 & $-$467$\pm$51 & $+$094$\pm$10 & 490$\pm$056& $-$1.92& 2.95 & 2.96 &  0.00$\pm$0.00 & 0.00$\pm$0.00& 1.00$\pm$0.00&BHB\\
 2353  & $-$004$\pm$54 & $-$037$\pm$54 & $+$160$\pm$11 & 164$\pm$077& $-$1.99& 2.94 & 2.95 &  0.00$\pm$0.00 & 0.01$\pm$0.00& 0.99$\pm$0.00&BHB\\
 2372  & $+$129$\pm$17 & $-$184$\pm$19 & $-$092$\pm$10 & 243$\pm$027& $-$1.49& 2.95 & 2.95 &  0.00$\pm$0.00 & 0.00$\pm$0.00& 1.00$\pm$0.00&BHB\\
 2376  & $-$138$\pm$16 & $-$022$\pm$15 & $+$116$\pm$12 & 182$\pm$025& $-$1.67& 2.23 & 2.91 &  0.00$\pm$0.00 & 0.02$\pm$0.00& 0.98$\pm$0.00&BHB\\
 2436  & $+$216$\pm$20 & $-$177$\pm$20 & $-$094$\pm$12 & 295$\pm$031& $-$1.43& 2.77 & 2.88 &  0.00$\pm$0.00 & 0.00$\pm$0.00& 1.00$\pm$0.00&BHB\\
 2441  & $-$168$\pm$52 & $-$041$\pm$38 & $-$050$\pm$15 & 180$\pm$066& $-$1.95& 2.71 & 2.98 &  0.01$\pm$0.25 & 0.09$\pm$0.20& 0.90$\pm$0.45&BHB\\
 2541  & $-$158$\pm$20 & $-$167$\pm$18 & $-$005$\pm$11 & 230$\pm$029& $-$1.61& 2.76 & 2.96 &  0.00$\pm$0.00 & 0.00$\pm$0.02& 1.00$\pm$0.02&BHB\\
 2585  & $+$023$\pm$51 & $-$329$\pm$48 & $-$062$\pm$34 & 336$\pm$078& $-$1.59& 2.23 & 2.76 &  0.00$\pm$0.00 & 0.00$\pm$0.00& 1.00$\pm$0.00&BHB\\
 1423  & $-$160$\pm$65 & $-$166$\pm$60 & $+$177$\pm$29 & 291$\pm$093& $-$1.60& 2.62 & 2.91 &  0.00$\pm$0.00 & 0.00$\pm$0.00& 1.00$\pm$0.00&BHB\\
 2713  & $+$007$\pm$50 & $-$324$\pm$48 & $+$219$\pm$36 & 391$\pm$078& $-$2.72& 2.24 & 2.80 &  0.00$\pm$0.00 & 0.00$\pm$0.00& 1.00$\pm$0.00&BHB\\
 2739  & $-$044$\pm$26 & $-$150$\pm$24 & $+$165$\pm$25 & 227$\pm$043& $-$1.55& 1.86 & 2.81 &  0.00$\pm$0.00 & 0.00$\pm$0.00& 1.00$\pm$0.00&BHB\\
 2758  & $+$167$\pm$17 & $-$240$\pm$14 & $-$096$\pm$13 & 308$\pm$026& $-$1.88& 2.10 & 2.80 &  0.00$\pm$0.00 & 0.00$\pm$0.00& 1.00$\pm$0.00&BHB\\
 2793  & $+$077$\pm$39 & $-$291$\pm$25 & $-$039$\pm$24 & 304$\pm$052& $-$1.60& 2.10 & 2.89 &  0.00$\pm$0.00 & 0.00$\pm$0.00& 1.00$\pm$0.00&BHB\\
 3077  & $-$107$\pm$82 & $-$166$\pm$58 & $+$050$\pm$58 & 204$\pm$116& $-$2.02& 2.07 & 2.94 &  0.00$\pm$0.00 & 0.04$\pm$0.08& 0.96$\pm$0.08&BHB\\
 3139  & $+$051$\pm$26 & $-$151$\pm$23 & $-$014$\pm$18 & 160$\pm$039& $-$1.50& 2.27 & 2.77 &  0.00$\pm$0.00 & 0.55$\pm$0.18& 0.45$\pm$0.18&BHB\\
 1618  & $+$049$\pm$24 & $-$426$\pm$30 & $+$059$\pm$18 & 433$\pm$042& $-$1.81& 2.53 & 2.95 &  0.00$\pm$0.00 & 0.00$\pm$0.00& 1.00$\pm$0.00&BHB\\
 2991  & $+$114$\pm$31 & $-$121$\pm$23 & $+$016$\pm$24 & 167$\pm$045& $-$1.42& 2.00 & 2.87 &  0.00$\pm$0.00 & 0.34$\pm$0.14& 0.66$\pm$0.14&BHB\\
         &      &    &    &      &       &   &     &    &    &     &    \\
Sample CA&      &    &    &      &       &   &     &    &    &     &    \\
 3148  & $+$084$\pm$19 & $-$016$\pm$16 & $+$055$\pm$14 & 102$\pm$29 & $-$0.86& 1.12 & 1.27 &  0.17$\pm$0.10 & 0.75$\pm$0.02& 0.08$\pm$0.13&A\\
 3028  & $-$029$\pm$39 & $-$091$\pm$32 & $-$089$\pm$33 & 131$\pm$60 & $-$0.56& 0.94 & 1.44 &  0.00$\pm$0.01 & 0.58$\pm$0.46& 0.42$\pm$0.47&A\\
 3033  & $+$047$\pm$13 & $-$021$\pm$11 & $-$046$\pm$11 & 069$\pm$20 & $-$1.60& 0.86 & 1.25 &  0.30$\pm$0.24 & 0.68$\pm$0.22& 0.02$\pm$0.01&A\\
 3046  & $+$100$\pm$18 & $-$031$\pm$12 & $-$010$\pm$11 & 105$\pm$24 & $-$0.28& 1.11 & 1.75 &  0.54$\pm$0.17 & 0.42$\pm$0.17& 0.04$\pm$0.01&A\\
 1628  & $+$019$\pm$48 & $-$072$\pm$45 & $-$043$\pm$23 & 086$\pm$70 & $+$0.19& 1.75 & 1.96 &  0.01$\pm$0.28 & 0.94$\pm$0.10& 0.05$\pm$0.19&A\\
 3216  & $+$129$\pm$38 & $-$135$\pm$36 & $-$077$\pm$26 & 202$\pm$58 & $+$0.00& 1.46 & 1.74 &  0.00$\pm$0.00 & 0.02$\pm$0.01& 0.98$\pm$0.01&A\\
 3358  & $+$024$\pm$09 & $+$012$\pm$10 & $+$001$\pm$11 & 027$\pm$17 & $-$1.28& 0.64 & 1.11 &  0.85$\pm$0.00 & 0.15$\pm$0.00& 0.00$\pm$0.00&A\\
 3180  & $+$126$\pm$34 & $-$035$\pm$31 & $-$024$\pm$22 & 133$\pm$51 & $-$1.10& 1.11 & 1.28 &  0.31$\pm$0.23 & 0.51$\pm$0.30& 0.18$\pm$0.07&A\\
 3361  & $+$039$\pm$09 & $-$039$\pm$10 & $-$019$\pm$11 & 058$\pm$17 & $+$0.00& 1.12 & 1.80 &  0.43$\pm$0.22 & 0.56$\pm$0.21& 0.01$\pm$0.01&A\\
 3867  & $+$118$\pm$44 & $-$140$\pm$47 & $-$035$\pm$23 & 186$\pm$68 & $-$1.49& 1.14 & 1.23 &  0.00$\pm$0.00 & 0.12$\pm$0.02& 0.88$\pm$0.02&A\\
 3379  & $+$059$\pm$31 & $-$014$\pm$32 & $+$043$\pm$30 & 074$\pm$54 & $-$0.25& 0.95 & 1.51 &  0.40$\pm$0.16 & 0.58$\pm$0.05& 0.02$\pm$0.11&A\\
 3382  & $-$036$\pm$12 & $-$011$\pm$12 & $-$012$\pm$10 & 040$\pm$20 & $-$0.31& 1.00 & 1.50 &  0.77$\pm$0.11 & 0.23$\pm$0.11& 0.00$\pm$0.00&A\\
 3384  & $+$076$\pm$11 & $+$005$\pm$11 & $+$034$\pm$11 & 083$\pm$19 & $-$1.21& 0.75 & 1.21 &  0.63$\pm$0.14 & 0.35$\pm$0.12& 0.02$\pm$0.02&A\\
 3394  & $-$028$\pm$14 & $-$025$\pm$14 & $+$008$\pm$13 & 038$\pm$24 & $-$1.26& 0.63 & 1.09 &  0.68$\pm$0.11 & 0.32$\pm$0.11& 0.00$\pm$0.00&A\\
 3272  & $+$129$\pm$20 & $-$038$\pm$20 & $+$041$\pm$17 & 141$\pm$33 & $-$0.57& 1.07 & 1.42 &  0.11$\pm$0.03 & 0.54$\pm$0.25& 0.35$\pm$0.28&A\\
 3438  & $+$083$\pm$13 & $-$084$\pm$20 & $+$018$\pm$22 & 119$\pm$32 & $-$0.11& 1.02 & 1.70 &  0.01$\pm$0.01 & 0.86$\pm$0.03& 0.13$\pm$0.02&A\\
 3455  & $+$136$\pm$41 & $-$008$\pm$50 & $-$058$\pm$46 & 148$\pm$79 & $+$0.00& 1.06 & 1.80 &  0.07$\pm$0.07 & 0.41$\pm$0.03& 0.52$\pm$0.05&A\\
 3904  & $+$151$\pm$23 & $+$012$\pm$23 & $+$050$\pm$17 & 160$\pm$37 & $-$0.40& 1.22 & 1.46 &  0.11$\pm$0.22 & 0.28$\pm$0.18& 0.61$\pm$0.40&A\\
 3468  & $+$040$\pm$13 & $+$013$\pm$15 & $-$009$\pm$14 & 043$\pm$24 & $-$0.21& 0.93 & 1.57 &  0.83$\pm$0.03 & 0.17$\pm$0.03& 0.00$\pm$0.00&A\\
 3916  & $+$264$\pm$22 & $-$244$\pm$35 & $+$128$\pm$20 & 382$\pm$46 & $-$1.01& 1.05 & 1.31 &  0.00$\pm$0.00 & 0.00$\pm$0.00& 1.00$\pm$0.00&A\\
 3474  & $+$046$\pm$26 & $-$073$\pm$35 & $+$053$\pm$30 & 101$\pm$53 & $-$1.04& 0.72 & 1.20 &  0.01$\pm$0.00 & 0.91$\pm$0.10& 0.09$\pm$0.10&A\\
 3920  & $-$165$\pm$38 & $-$067$\pm$45 & $-$052$\pm$28 & 186$\pm$65 & $+$0.00& 1.38 & 1.68 &  0.00$\pm$0.23 & 0.06$\pm$0.20& 0.94$\pm$0.42&A\\
 3561  & $-$020$\pm$14 & $-$022$\pm$13 & $-$026$\pm$13 & 039$\pm$23 & $-$1.29& 0.74 & 1.13 &  0.60$\pm$0.26 & 0.40$\pm$0.25& 0.00$\pm$0.01&A\\
 3564  & $+$015$\pm$19 & $+$016$\pm$22 & $-$010$\pm$19 & 024$\pm$35 & $+$0.00& 1.15 & 1.67 &  0.84$\pm$0.05 & 0.16$\pm$0.05& 0.00$\pm$0.00&A\\
 1674  & $+$054$\pm$31 & $-$036$\pm$40 & $-$085$\pm$28 & 107$\pm$58 & $-$0.88& 1.14 & 1.56 &  0.00$\pm$0.10 & 0.80$\pm$0.22& 0.20$\pm$0.33&A\\
 3661  & $-$218$\pm$34 & $-$438$\pm$70 & $-$001$\pm$39 & 489$\pm$87 & $+$0.00& 1.16 & 1.83 &  0.00$\pm$0.00 & 0.00$\pm$0.00& 1.00$\pm$0.00&A\\
 3665  & $+$057$\pm$31 & $-$081$\pm$49 & $-$052$\pm$39 & 112$\pm$70 & $-$0.07& 1.01 & 1.65 &  0.00$\pm$0.26 & 0.87$\pm$0.12& 0.13$\pm$0.39&A\\
 3951  & $+$029$\pm$19 & $+$050$\pm$21 & $+$009$\pm$16 & 058$\pm$33 & $+$0.00& 1.31 & 1.82 &  0.69$\pm$0.26 & 0.31$\pm$0.25& 0.00$\pm$0.01&A\\
 3960  & $+$197$\pm$24 & $-$108$\pm$28 & $-$009$\pm$21 & 225$\pm$42 & $-$0.10& 1.21 & 1.69 &  0.00$\pm$0.00 & 0.00$\pm$0.00& 1.00$\pm$0.00&A\\
 3758  & $-$116$\pm$24 & $+$000$\pm$33 & $+$022$\pm$29 & 118$\pm$50 & $-$1.06& 0.81 & 1.31 &  0.68$\pm$0.04 & 0.28$\pm$0.14& 0.04$\pm$0.09&A\\
 3992  & $-$035$\pm$18 & $-$066$\pm$23 & $+$000$\pm$19 & 075$\pm$35 & $-$0.38& 0.94 & 1.52 &  0.11$\pm$0.18 & 0.87$\pm$0.15& 0.02$\pm$0.03&A\\
 4009  & $+$109$\pm$38 & $-$040$\pm$54 & $-$056$\pm$47 & 129$\pm$81 & $-$0.10& 0.89 & 1.54 &  0.04$\pm$0.32 & 0.68$\pm$0.03& 0.28$\pm$0.34&A\\
 4013  & $+$047$\pm$20 & $-$047$\pm$25 & $-$006$\pm$22 & 067$\pm$39 & $+$0.00& 1.09 & 1.82 &  0.36$\pm$0.31 & 0.63$\pm$0.30& 0.01$\pm$0.01&A\\
 1784  & $+$004$\pm$22 & $+$070$\pm$23 & $+$038$\pm$13 & 080$\pm$34 & $-$0.17& 1.43 & 1.57 &  0.19$\pm$0.30 & 0.79$\pm$0.27& 0.02$\pm$0.03&A\\
 1859  & $+$082$\pm$14 & $-$114$\pm$18 & $+$044$\pm$10 & 147$\pm$25 & $-$0.30& 1.57 & 1.68 &  0.00$\pm$0.00 & 0.59$\pm$0.02& 0.41$\pm$0.02&A\\
 1927  & $-$163$\pm$21 & $+$016$\pm$22 & $-$043$\pm$24 & 169$\pm$39 & $+$0.38& 1.24 & 1.90 &  0.12$\pm$0.25 & 0.18$\pm$0.13& 0.70$\pm$0.38&A\\
 2050  & $-$087$\pm$17 & $-$009$\pm$12 & $+$022$\pm$10 & 090$\pm$23 & $-$0.32& 1.50 & 1.68 &  0.69$\pm$0.01 & 0.30$\pm$0.00& 0.01$\pm$0.01&A\\
 2084  & $-$039$\pm$38 & $-$039$\pm$35 & $+$040$\pm$19 & 068$\pm$55 & $-$0.95& 1.20 & 1.31 &  0.21$\pm$0.08 & 0.77$\pm$0.05& 0.02$\pm$0.04&A\\
 2158  & $+$061$\pm$08 & $+$031$\pm$07 & $+$021$\pm$10 & 072$\pm$15 & $-$1.63& 1.21 & 1.24 &  0.74$\pm$0.11 & 0.25$\pm$0.10& 0.01$\pm$0.01&A\\
 2204  & $-$075$\pm$11 & $-$042$\pm$07 & $+$159$\pm$10 & 181$\pm$16 & $+$0.00& 1.72 & 1.74 &  0.00$\pm$0.00 & 0.00$\pm$0.00& 1.00$\pm$0.00&A\\
 2206  & $+$087$\pm$14 & $+$021$\pm$14 & $+$040$\pm$15 & 098$\pm$25 & $+$0.00& 1.27 & 1.67 &  0.53$\pm$0.28 & 0.44$\pm$0.21& 0.03$\pm$0.07&A\\
 2258  & $-$010$\pm$16 & $-$127$\pm$23 & $-$040$\pm$12 & 134$\pm$30 & $-$0.42& 1.57 & 1.60 &  0.00$\pm$0.00 & 0.79$\pm$0.26& 0.21$\pm$0.26&A\\
 2435  & $+$078$\pm$10 & $-$022$\pm$07 & $+$050$\pm$09 & 095$\pm$15 & $-$1.35& 1.23 & 1.26 &  0.21$\pm$0.08 & 0.74$\pm$0.04& 0.06$\pm$0.05&A\\
 2579  & $-$074$\pm$20 & $+$008$\pm$16 & $+$045$\pm$14 & 087$\pm$29 & $-$1.46& 1.02 & 1.23 &  0.46$\pm$0.20 & 0.52$\pm$0.19& 0.02$\pm$0.01&A\\
 2868  & $+$018$\pm$27 & $+$053$\pm$10 & $-$006$\pm$11 & 056$\pm$31 & $-$0.39& 0.89 & 1.44 &  0.68$\pm$0.09 & 0.32$\pm$0.09& 0.00$\pm$0.00&A\\
 2886  & $-$015$\pm$08 & $-$020$\pm$09 & $-$018$\pm$08 & 031$\pm$14 & $-$0.32& 0.91 & 1.57 &  0.68$\pm$0.13 & 0.32$\pm$0.13& 0.00$\pm$0.00&A\\
 2893  & $+$015$\pm$17 & $-$006$\pm$12 & $-$002$\pm$14 & 016$\pm$25 & $-$0.93& 0.86 & 1.47 &  0.82$\pm$0.06 & 0.18$\pm$0.05& 0.00$\pm$0.00&A\\
 2908  & $+$005$\pm$31 & $-$014$\pm$21 & $+$059$\pm$26 & 061$\pm$46 & $-$0.53& 0.71 & 1.18 &  0.16$\pm$0.17 & 0.82$\pm$0.11& 0.02$\pm$0.06&A\\
 3080  & $+$008$\pm$13 & $-$084$\pm$19 & $-$141$\pm$19 & 164$\pm$30 & $+$0.00& 1.22 & 1.68 &  0.00$\pm$0.00 & 0.03$\pm$0.09& 0.97$\pm$0.09&A\\
 3103  & $-$010$\pm$12 & $-$018$\pm$12 & $-$022$\pm$11 & 030$\pm$20 & $-$0.55& 1.09 & 1.43 &  0.67$\pm$0.19 & 0.33$\pm$0.19& 0.00$\pm$0.00&A\\
\enddata

\tablenotetext{a}{  Total Space Velocity (km/s).} 
\tablenotetext{b}{  Height of star above Galactic plane (kpc).} 
\tablenotetext{c}{ Distance of star (kpc).} 
\tablenotetext{d}{ Probability that star belongs to thin disk. } 
\tablenotetext{e}{ Probability that star belongs to thick disk. } 
\tablenotetext{f}{ Probability that star belongs to the halo. } 

\end{deluxetable}

\end{document}